\documentclass[ALICE,manyauthors]{cernphprep}
\usepackage[comma,square,numbers,sort&compress]{natbib}
\usepackage{hyperref}
\usepackage{lineno}
\usepackage{xcolor}
\usepackage{xspace}
\usepackage{multirow}
\usepackage{makecell}
\usepackage{booktabs}
\usepackage{upgreek}
\usepackage[T1]{fontenc}

\begin{document}
%

\newcommand{\ee}               {\mathrm{e^+e^-}} 
\newcommand{\pp}               {\mathrm{pp}}
\newcommand{\ppbar}            {\mathrm{p\overline{p}}}
\newcommand{\pPb}              {\mathrm{p--Pb}}
\newcommand{\PbPb}             {\mathrm{Pb--Pb}}

\newcommand{\av}[1]            {\left\langle #1 \right\rangle}

\newcommand{\s}                {\sqrt{s}}
\newcommand{\sqrtsNN}          {\sqrt{s_\mathrm{NN}}}
\newcommand{\Npart}            {N_\mathrm{part}}
\newcommand{\Ncoll}            {N_\mathrm{coll}}
\newcommand{\RpPb}             {R_\mathrm{pPb}}
\newcommand{\RAA}              {R_\mathrm{AA}}
\newcommand{\TAA}              {T_\mathrm{AA}}
\newcommand{\pt}               {p_\mathrm{T}}
\newcommand{\mt}               {m_\mathrm{T}}
\newcommand{\de}               {\mathrm{d}}
\newcommand{\dEdx}             {\de E/\de x}
\newcommand{\dNdpt}            {\de N/\de\pt}
\newcommand{\dNdy}             {\de N/\de y}
\newcommand{\mur}              {\mu_\mathrm{R}}
\newcommand{\muf}              {\mu_\mathrm{F}}
\newcommand{\T}                {\mathrm{T}}
\newcommand{\meanpt}           {\langle p_\mathrm{T} \rangle}
\newcommand{\sigmatot}         {\sigma_{\rm tot}}
\newcommand{\fprompt}          {f_\mathrm{prompt}}
\newcommand{\fnonprompt}       {f_\mathrm{non\text{-}prompt}}
\newcommand{\f}[1]             {f_\mathrm{#1}}
\newcommand{\muR}              {\mu_\mathrm{R}}
\newcommand{\muF}              {\mu_\mathrm{F}}

\newcommand{\eV}               {\mathrm{eV}}
\newcommand{\keV}              {\mathrm{keV}}
\newcommand{\MeV}              {\mathrm{MeV}}
\newcommand{\GeV}              {\mathrm{GeV}}
\newcommand{\TeV}              {\mathrm{TeV}}
\newcommand{\ev}               {\mathrm{eV}}
\newcommand{\kev}              {\mathrm{keV}}
\newcommand{\mev}              {\mathrm{MeV}}
\newcommand{\mevc}             {\mathrm{MeV}/c}
\newcommand{\mevcsquared}      {\mathrm{MeV}/c^2}
\newcommand{\gev}              {\mathrm{GeV}}
\newcommand{\gevc}             {\mathrm{GeV}/c}
\newcommand{\tev}              {\mathrm{TeV}}
\newcommand{\fm}               {\mathrm{fm}}
\newcommand{\mm}               {\mathrm{mm}} 
\newcommand{\cm}               {\mathrm{cm}}
\newcommand{\m}                {\mathrm{m}}
\newcommand{\mum}              {\mathrm{\upmu m}}
\newcommand{\ns}               {\mathrm{ns}}
\newcommand{\mrad}             {\mathrm{mrad}}
\newcommand{\mb}               {\mathrm{mb}}
\newcommand{\mub}              {\mathrm{\upmu b}}
\newcommand{\lumi}             {\mathcal{L}_\mathrm{int}}
\newcommand{\nbinv}            {\mathrm{nb^{-1}}}

\newcommand{\ITS}              {\mathrm{ITS}}
\newcommand{\TOF}              {\mathrm{TOF}}
\newcommand{\ZDC}              {\mathrm{ZDC}}
\newcommand{\ZDCs}             {\mathrm{ZDC}}
\newcommand{\ZNA}              {\mathrm{ZNA}}
\newcommand{\ZNC}              {\mathrm{ZNC}}
\newcommand{\SPD}              {\mathrm{SPD}}
\newcommand{\SDD}              {\mathrm{SDD}}
\newcommand{\SSD}              {\mathrm{SSD}}
\newcommand{\TPC}              {\mathrm{TPC}}
\newcommand{\TRD}              {\mathrm{TRD}}
\newcommand{\VZERO}            {\mathrm{V0}}
\newcommand{\VZEROA}           {\mathrm{V0A}}
\newcommand{\VZEROC}           {\mathrm{V0C}}

\newcommand{\pip}              {\mathrm{\uppi^{+}}}
\newcommand{\pim}              {\mathrm{\uppi^{-}}}
\newcommand{\kap}              {\mathrm{\rm{K}^{+}}}
\newcommand{\kam}              {\mathrm{\rm{K}^{-}}}
\newcommand{\pbar}             {\mathrm{\rm\overline{p}}}
\newcommand{\kzero}            {\mathrm{K}^0_S}
\newcommand{\lmb}              {\mathrm{\Lambda}}
\newcommand{\almb}             {\mathrm{\overline{\Lambda}}}
\newcommand{\Om}               {\mathrm{\Omega^-}}
\newcommand{\Mo}               {\mathrm{\overline{\Omega}^+}}
\newcommand{\X}                {\mathrm{\Xi^-}}
\newcommand{\Ix}               {\mathrm{\overline{\Xi}^+}}
\newcommand{\Xis}              {\mathrm{\Xi^{\pm}}}
\newcommand{\Oms}              {\mathrm{\Omega^{\pm}}}
\newcommand{\DzerotoKpi}       {\mathrm{D^0 \to K^-\uppi^+}}
\newcommand{\DplustoKpipi}     {\mathrm{D^+\to K^-\uppi^+\uppi^+}}
\newcommand{\DstartoDpi}       {\mathrm{D^{*+} \to \rm D^0 \uppi^+}}
\newcommand{\Dstophipi}        {\mathrm{D_s^+\to \upphi\uppi^+}}
\newcommand{\Dstophipipm}      {\mathrm{D_s^\pm\to \upphi\uppi^\pm}}
\newcommand{\DstophipitoKKpi}  {\mathrm{D_s^+\to \upphi\uppi^+\to K^-K^+\uppi^+}}
\newcommand{\DplustoKKpi}      {\mathrm{D^+\to K^-K^+\uppi^+}}
\newcommand{\phitoKK}          {\mathrm{\upphi\to  K^-K^+}}
\newcommand{\DstoKzerostarK}   {\mathrm{D_s^+\to \overline{K}^{*0} K^+}}
\newcommand{\Dstofzeropi}      {\mathrm{D_s^+\to f_0(980) \uppi^+}}
\newcommand{\fzero}            {\mathrm{f_0(980)}}
\newcommand{\Kzerostar}        {\mathrm{\overline{K}^{*0}}}
\newcommand{\Dzero}            {\mathrm{D^0}}
\newcommand{\Dzerobar}         {\mathrm{\overline{D}\,^0}}
\newcommand{\Dstar}            {\mathrm{D^{*+}}}
\newcommand{\Dstarm}           {\mathrm{D^{*-}}}
\newcommand{\DstarZero}        {\mathrm{D^{*0}}}
\newcommand{\DstarS}           {\mathrm{D_s^{*+}}}
\newcommand{\Dplus}            {\mathrm{D^+}}
\newcommand{\Dminus}           {\mathrm{D^-}}
\newcommand{\Ds}               {\mathrm{D_s^+}}
\newcommand{\Dspm}             {\mathrm{D_s^\pm}}
\newcommand{\Dsstar}           {\mathrm{D_s^{*+}}}
\newcommand{\KKpi}             {\mathrm{K^-K^+\uppi^+}}
\newcommand{\cubar}            {\mathrm{c\bar{u}}}
\newcommand{\cdbar}            {\mathrm{c\bar{d}}}
\newcommand{\ccbar}            {\mathrm{c\overline{c}}}
\newcommand{\bbbar}            {\mathrm{b\overline{b}}}
\newcommand{\Bzero}            {\mathrm{B^0}}
\newcommand{\Bplus}            {\mathrm{B^+}}
\newcommand{\Bzeroplus}        {\mathrm{B^{0,+}}}
\newcommand{\Bs}               {\mathrm{B_s^0}}
\newcommand{\Lambdab}          {\mathrm{\Lambda_b^0}}
\newcommand{\Jpsi}             {\mathrm{J}/\uppsi}
\newcommand{\Vdecay} 	       {\mathrm{V^{0}}}
\newcommand{\bhad}             {\mathrm{H_b}}
\newcommand{\Ztobbbar}         {\mathrm{Z\to b\overline{b}}}
\newcommand{\fctoD}            {f(\mathrm{c}\to\mathrm{D})}
\newcommand{\fbtoB}            {f(\mathrm{b}\to\mathrm{B})}
\newcommand{\fctoHc}           {f(\mathrm{c}\to\mathrm{H_c})}
\newcommand{\fbtoHb}           {f(\mathrm{b}\to\mathrm{H_b})}
\newcommand{\btoDX}            {\mathrm{b\to D+X}}
\newcommand{\rawY}[1]          {Y_{#1}}
\newcommand{\effNP}[1]         {(\mathrm{Acc}\times\epsilon)^\mathrm{non\text{-}prompt}_{#1}}
\newcommand{\effP}[1]          {(\mathrm{Acc}\times\epsilon)^\mathrm{prompt}_{#1}}
\newcommand{\Np}               {N_\mathrm{prompt}}
\newcommand{\Nnp}              {N_\mathrm{non\text{-}prompt}}

\begin{titlepage}
\PHyear{2021}       
\PHnumber{034}      
\PHdate{23 February}  

\title{Measurement of beauty and charm production in pp collisions at $\pmb{\s=5.02~\TeV}$ via non-prompt and prompt D mesons}
\ShortTitle{Non-prompt and prompt D-meson production in pp collisions at $\s=5.02~\TeV$}   

\Collaboration{ALICE Collaboration\thanks{See Appendix~\ref{app:collab} for the list of collaboration members}}
\ShortAuthor{ALICE Collaboration} 

\begin{abstract}
The $\pt$-differential production cross sections of prompt and non-prompt (produced in beauty-hadron decays) D mesons were measured by the ALICE experiment at midrapidity ($|y|<0.5$) in proton--proton collisions at $\s=5.02~\TeV$. The data sample used in the analysis corresponds to an integrated luminosity of $(19.3\pm0.4)~\nbinv$. D mesons were reconstructed from their decays $\DzerotoKpi$, $\DplustoKpipi$, and $\DstophipitoKKpi$ and their charge conjugates. Compared to previous measurements in the same rapidity region, the cross sections of prompt $\Dplus$ and $\Ds$ mesons have an extended $\pt$ coverage and total uncertainties reduced by a factor ranging from 1.05 to 1.6, depending on $\pt$, allowing for a more precise determination of their $\pt$-integrated cross sections. The results are well described by perturbative QCD calculations. The fragmentation fraction of heavy quarks to strange mesons divided by the one to non-strange mesons, $\f{s}/(\f{u}+\f{d})$, is compatible for charm and beauty quarks and with previous measurements at different centre-of-mass energies and collision systems. The $\bbbar$ production cross section per rapidity unit at midrapidity, estimated from non-prompt D-meson measurements, is $\de\sigma_\bbbar/\de y|_\mathrm{|y|<0.5} = 34.5 \pm 2.4 (\mathrm{stat}) ^{+4.7}_{-2.9} (\mathrm{tot.~syst})~\mub$. It is compatible with previous measurements at the same centre-of-mass energy and with the cross section predicted by perturbative QCD calculations.

\end{abstract}
\end{titlepage}

\setcounter{page}{2} 

\section{Introduction}
\label{sec:introduction}
Measurements of the production of hadrons containing charm or beauty quarks in proton--proton (pp) collisions provide an important test of Quantum Chromodynamics (QCD) calculations.  
They also set the reference for the respective measurements in heavy-ion collisions, where the study of charm- and beauty-quark interaction with the quark--gluon plasma (QGP) constituents is a rich source of information about the medium properties and its inner dynamics~\cite{Prino:2016cni}.    
Several measurements of charm and beauty production were carried out in pp collisions at $\s=2.76,~5.02,~7,~8$, and 13 TeV by the ALICE~\cite{Abelev:2012vra,Acharya:2019mgn,Acharya:2017jgo,Abelev:2012tca,Acharya:2017kfy,Abelev:2012qh,Abelev:2012pi,Abelev:2014hla,Abelev:2012gx,Acharya:2018ohw,Abelev:2012sca,Acharya:2018kkj,Acharya:2019mom,Acharya:2019mky,Acharya:2017lwf}, ATLAS~\cite{Aad:2012jga,Aad:2015zix,ATLAS:2013cia,Aad:2015cda,Aaboud:2019wtr}, CMS~\cite{Chatrchyan:2012hw,Sirunyan:2017xss,Sirunyan:2019fnc,Khachatryan:2010yr,Khachatryan:2011mk,Chatrchyan:2011pw,Chatrchyan:2011vh,Khachatryan:2016csy,Sirunyan:2018ktu,Khachatryan:2014nfa}, and LHCb~\cite{Aaij:2010gn,Aaij:2016jht,Aaij:2013mga,Aaij:2015bpa,Aaij:2012jd,Aaij:2011jp,Aaij:2012dd,Aaij:2013noa,Aaij:2014ija,Aaij:2015fea,Aaij:2016avz,Aaij:2019ezy,Aaij:2019pqz,Aaij:2019zxa,Aaij:2019ths} experiments at the LHC. At lower collision energies, measurements were performed at $\s = 200~\GeV$ at RHIC~\cite{Adare:2006hc,Adare:2009ic,Adamczyk:2012af,Adare:2017caq} and in $\ppbar$ collisions at $\s = 630~\GeV$ at the S$\ppbar$S~\cite{Albajar:1990zu} and at $\s = 1.96~\TeV$ at  the Tevatron~\cite{Acosta:2003ax,Acosta:2004yw,Abulencia:2006ps,Aaltonen:2009xn}. 
The D- and B-meson data are generally described within uncertainties by perturbative QCD calculations at Next-to-Leading-Order with Next-to-Leading Log resummation, like FONLL~\cite{Cacciari:1998it,Cacciari:2001td} and GM-VFNS~\cite{Kniehl:2004fy,Kniehl:2012ti,Benzke:2017yjn,Kramer:2017gct,Helenius:2018uul,Bolzoni:2013vya}. These calculations rely on the factorisation of soft (non-perturbative) and hard (perturbative) processes and calculate the transverse-momentum $(\pt)$ differential cross sections of charm- or beauty-hadron production as a convolution of a hard-scattering cross section at the partonic level, parton distribution functions (PDFs) of the colliding protons, and  fragmentation functions (FF) modelling the transition from heavy quarks to heavy-flavour hadrons~\cite{Collins:1989gx}. Recently, also calculations with next-to-next-to-leading-order (NNLO) QCD radiative corrections became available for the beauty-quark production~\cite{Catani:2020kkl}.

In this paper we report an update of the measurement of prompt (i.e.\ produced in the charm quark fragmentation, either directly or through decays of excited open charm and charmonium states) $\Dplus$- and $\Ds$-meson production performed with ALICE in the rapidity interval $|y|<0.5$ in pp collisions at $\sqrt{s}=5.02~\tev$~\cite{Acharya:2019mgn}, obtained using an improved analysis technique. We also present a new measurement of the production of non-prompt $\Dzero$, $\Dplus$, and $\Ds$ mesons from beauty-hadron decays. The analysis of prompt $\Dplus$ and $\Ds$ mesons is extended down to $\pt=0$ and $1~\gevc$, respectively. Non-prompt D mesons are measured down to $\pt=1~\gevc$ ($\Dzero$ meson) and $2~\gevc$ ($\Dplus$ and $\Ds$ mesons). These new results provide an improvement in terms of low-$\pt$ reach and particle species accessed with respect to the previous measurement of non-prompt $\Dzero$ production by CMS~\cite{Sirunyan:2018ktu}. Such an extension is important to test perturbative QCD (pQCD) calculations over a wider $\pt$ interval and to better determine the heavy-quark production cross section. These measurements also provide a reference for Pb--Pb collisions in the low-$\pt$ region, a relevant one to address nuclear effects like shadowing, heavy-quark diffusion in the QGP, and the expected enhancement of the production of hadrons with strange quarks~\cite{Rapp:2018qla}.

The paper is organised as follows. In Section~\ref{sec:apparatus} the ALICE apparatus and the analysed data sample are described. In Section~\ref{sec:analysis} the analysis procedure is explained. Machine-learning algorithms are used to classify and separate the prompt and non-prompt D-meson signals and the combinatorial background. A data-driven procedure is used to calculate the fraction of prompt and non-prompt D mesons. The systematic uncertainties are discussed in Section~\ref{sec:systematic}. In Section~\ref{sec:results} the results are presented. First, in Section~\ref{sec:NPD_cross_sec}, the $\pt$-differential cross sections of prompt and non-prompt D mesons are reported and compared to theoretical predictions. Then, in Section~\ref{sec:resultsRatios}, the ratios of the measured cross sections of the D-meson species are computed. In theoretical calculations, these ratios are sensitive mainly to the FF or the adopted hadronisation model. In particular, the comparison of the production rate of strange mesons with that of non-strange ones allows the determination of the ratio $\f{s}/(\f{u}+\f{d})$, i.e. the fragmentation fraction of charm and beauty quarks to strange mesons divided by the one to non-strange mesons. 
In Section~\ref{sec:resultsBeauty}, by extrapolating down to $\pt=0$ the measured non-prompt D-meson cross sections, an estimate of the production cross section of beauty quarks  at midrapidity is obtained, which represents the most-precise result to date in pp collisions at $\s=5.02~\tev$. A summary concludes the paper.

\section{Experimental apparatus and data sample}
\label{sec:apparatus}

The ALICE apparatus is composed of a central barrel, consisting
of a set of detectors for particle reconstruction and identification at
midrapidity, a forward muon spectrometer, and various forward and backward
detectors for triggering and event characterisation.
A complete description and an overview of their typical performance are
presented in Refs.~\cite{Aamodt:2008zz,Abelev:2014ffa}.

The D-meson decay products were reconstructed at midrapidity exploiting
the tracking and particle identification capabilities of the central barrel
detectors, which cover the full azimuth in the pseudorapidity interval
$|\eta| < 0.9$.
These detectors are embedded in a large solenoidal magnet that provides a
magnetic field $B = 0.5~\mathrm{T}$ parallel to the beam direction.
Charged-particle tracks are reconstructed from their hits in the
Inner Tracking System (ITS) and the Time Projection Chamber (TPC).
The ITS is the innermost ALICE detector; it consists of six cylindrical layers
of silicon detectors, allowing a precise determination of the track parameters
in the vicinity of the interaction point.
The TPC provides up to 159 three-dimensional space points to reconstruct the
charged-particle trajectory, as well as particle identification via the
measurement of the specific ionisation energy loss $\dEdx$.
The particle identification capabilities of the TPC are extended by 
the Time-Of-Flight (TOF) detector, which is used to measure the flight time
of the charged particles from the interaction point. The event collision time is obtained using either the information from the
T0 detector, the TOF detector, or a combination of the two.
The T0 detector consists of two arrays of {\v C}erenkov counters, located on both sides of the nominal interaction point, covering the pseudorapidity intervals
$-3.28 < \eta < -2.97$ and $4.61 < \eta < 4.92$.
The V0 detector was used for triggering and event selection.
It is composed of two scintillator arrays, located on both sides of the
nominal interaction point and covering the pseudorapidity intervals
$-3.7 < \eta < -1.7$ and $2.8 < \eta < 5.1$.

The results presented in this paper were obtained from the analysis of
the data sample of pp collisions at $\s = 5.02$~TeV collected in 2017.
The events used in the analysis were recorded with a minimum bias (MB)
trigger which required coincident signals in the two scintillator
arrays of the V0 detector.
Events were further selected offline in order to remove background due to
the interaction between one of the beams and the residual gas present in the
beam vacuum tube and other machine-induced backgrounds~\cite{Abelev:2014ffa}.
This selection was based on the timing information of the two V0 arrays
and the correlation between the number of hits and track segments 
in the two innermost layers of the ITS, consisting of Silicon Pixel Detectors (SPD).
In order to maintain a uniform acceptance in pseudorapidity, events were
required to have a reconstructed collision vertex located within
$\pm10~\rm{cm}$ from the centre of the detector along the beam-line direction.
Events with multiple primary vertices reconstructed from TPC and ITS tracks, due to pileup of several collisions, were rejected.
The rejected pileup events amount to about 1\% of the triggered events and the remaining undetected pileup is negligible in the present analysis. 
After the aforementioned selections, the data sample used for the analysis
consists of about 990 million MB events, corresponding to an integrated
luminosity $\lumi = (19.3 \pm 0.4)$~nb$^{-1}$~\cite{ALICE-PUBLIC-2018-014}.

The Monte Carlo samples utilised in the analysis were obtained simulating pp collisions with the PYTHIA~8.243 event generator~\cite{Sjostrand:2006za, Sjostrand:2014zea} (Monash-13 tune~\cite{Skands:2014pea}), and propagating the generated particles through the detector using the GEANT3 package~\cite{Brun:1994aa}. A $\ccbar$- or $\bbbar$-quark pair was required in each simulated PYTHIA pp event and D mesons were forced to decay into the hadronic channels of interest for the analysis. The luminous region distribution and the conditions of all the ALICE detectors in terms of active channels, gain, noise level, and alignment, and their evolution with time during the data taking, were taken into account in the simulations.
\section{Analysis technique}
\label{sec:analysis}
$\Dzero$, $\Dplus$, and $\Ds$ mesons and their charge conjugates were reconstructed through the decay channels $\DzerotoKpi$ (with branching ratio $\mathrm{BR} = (3.950 \pm 0.031) \%$), $\DplustoKpipi$ ($\mathrm{BR} = (9.38 \pm 0.16) \%$), and $\DstophipitoKKpi$ ($\mathrm{BR} = (2.24 \pm 0.08) \%$)~\cite{Zyla:2020zbs}. The analysis was based on the reconstruction of decay-vertex topologies displaced from the interaction vertex. The separation induced by the weak decays of prompt $\Dzero$, $\Dplus$, and $\Ds$ is typically a few hundred of $\mum$ ($c\tau \simeq 123$, $312$, and $151~\mum$, respectively~\cite{Zyla:2020zbs}). Decay vertices of non-prompt D mesons, originating from beauty-hadron decays, on average are more displaced from the interaction vertex due to the larger mean proper decay lengths of beauty hadrons ($c\tau \simeq 500~\mum$~\cite{Zyla:2020zbs})]) as compared to charm hadrons. Therefore, exploiting the selection of displaced decay-vertex topologies, it is possible not only to separate D mesons from the combinatorial background, but also non-prompt from prompt D mesons.

D-meson candidates were built combining pairs or triplets of tracks with the proper charge signs, each with $|\eta| < 0.8$, $\pt > 0.3~\gev/c$, at least 70 (out of 159) associated space points in the TPC, a fit quality $\chi^{2}/{\rm ndf} < 2$ in the TPC (where ndf is the number of degrees of freedom involved in the track fit procedure), and a minimum of two (out of six) hits in the ITS, with at least one in either of the two innermost SPD layers, which provide the best pointing resolution.
These track-selection criteria reduce the D-meson acceptance in rapidity, which drops steeply to zero for $|y|>0.5$ at low $\pt$ and for $|y|>0.8$ at $\pt>5~\gevc$. Thus, only D-meson candidates within a fiducial acceptance region, $|y| < y_{\rm fid}(\pt)$, were selected. The $y_{\mathrm{fid}}(\pt)$ value was defined as a second-order polynomial function, increasing from 0.5 to 0.8 in the transverse-momentum range $0 < \pt < 5~\gevc$, and as a constant term, $y_{\mathrm{fid}}=0.8$, for $\pt > 5~\gevc$.

To reduce the large combinatorial background and to separate the contribution of prompt and non-prompt D mesons, a machine-learning approach based on Boosted Decision Trees (BDT) was adopted. Two different implementations of the BDT algorithm, provided by the TMVA~\cite{Voss:2009rK} and XGBoost~\cite{Chen:2016XST} libraries, were considered. Signal samples of prompt and non-prompt D mesons for the BDT training were obtained from simulations based on the PYTHIA~8 event generator as described in Section~\ref{sec:apparatus}. The background samples were obtained from the sidebands of the candidate invariant-mass distributions in the data. Before the training, loose kinematic and topological selections were applied to the D-meson candidates together with the particle identification (PID) of decay-product tracks. Pions and kaons were selected by requiring compatibility with the respective particle hypothesis within three standard deviations ($3\,\sigma$) between the measured and the expected signals for both the TPC $\mathrm{d}E/\mathrm{d}x$ and the time of flight. Tracks without TOF hits were identified using only the TPC information. For $\Ds$-meson candidates, an absolute difference of the reconstructed ${\rm K^+K^-}$ invariant mass with respect to the PDG world average of the $\phi$ meson~\cite{Zyla:2020zbs} ($\Delta M_{\mathrm{KK}}$) under $15~\mevcsquared$ was additionally required.
The D-meson candidate information provided to the BDTs, as an input for the models to distinguish among prompt and non-prompt D mesons and background candidates, was mainly based on the displacement of the tracks from the primary vertex ($d_0$), the distance between the D-meson decay vertex and the primary vertex (decay length, $L$), the D-meson impact parameter, and the cosine of the pointing angle between the D-meson candidate line of flight (the vector connecting the primary and secondary vertices) and its reconstructed momentum vector. Additional variables related to the PID of decay tracks were used for $\Dplus$ and $\Ds$ candidates. The value of $\Delta M_{\mathrm{KK}}$ was also considered for $\Ds$ candidates. 
Independent BDTs were trained for the different D-meson species and in different $\pt$ intervals. Subsequently, they were applied to the real data sample in which the type of candidate is unknown. The BDT outputs are related to the candidate probability to be a non-prompt D meson or combinatorial background. Selections on the BDT outputs were optimised to obtain a high non-prompt D-meson fraction while maintaining a reliable signal extraction in the case of the non-prompt analysis. For the prompt $\Dplus$ and $\Ds$ analysis, selections were tuned to provide a large statistical significance for the signal and a small contribution of non-prompt candidates.

\subsection{Measurement of non-prompt $\Dzero$, $\Dplus$, and $\Ds$ mesons}
\label{sec:analysis_non_prompt}

\begin{figure}[tb]
    \begin{center}
    \includegraphics[width = 0.32\textwidth]{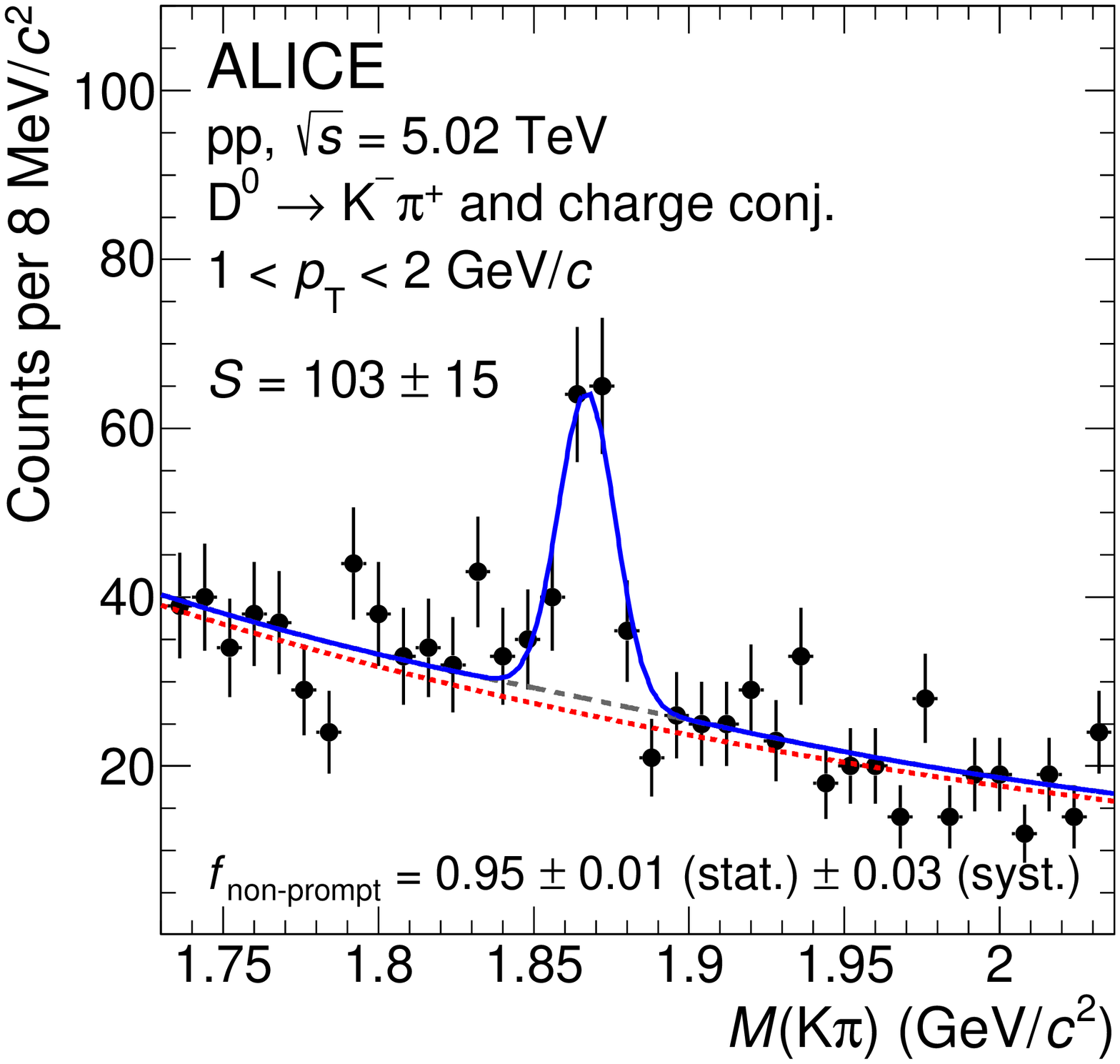}
    \includegraphics[width = 0.32\textwidth]{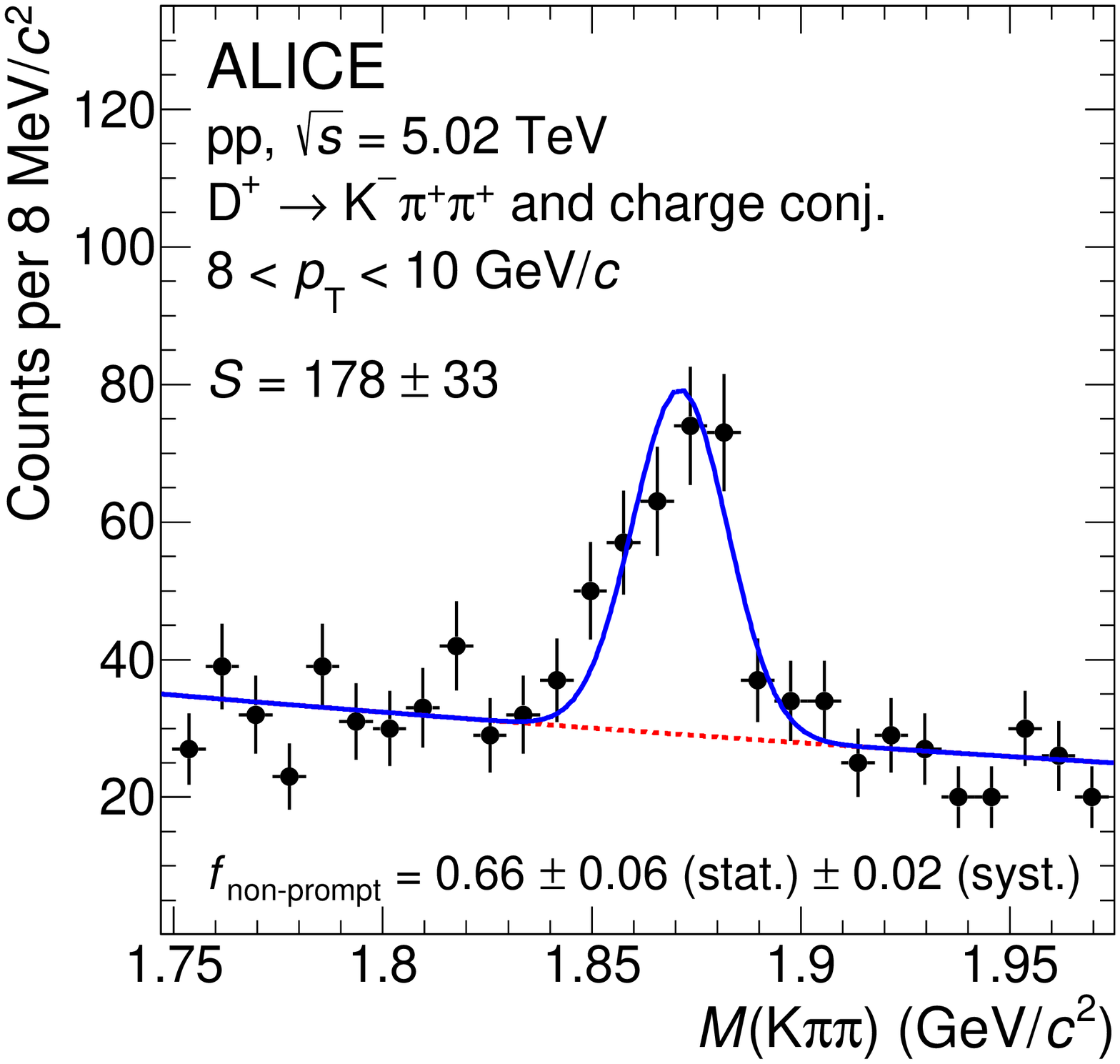}
    \includegraphics[width = 0.32\textwidth]{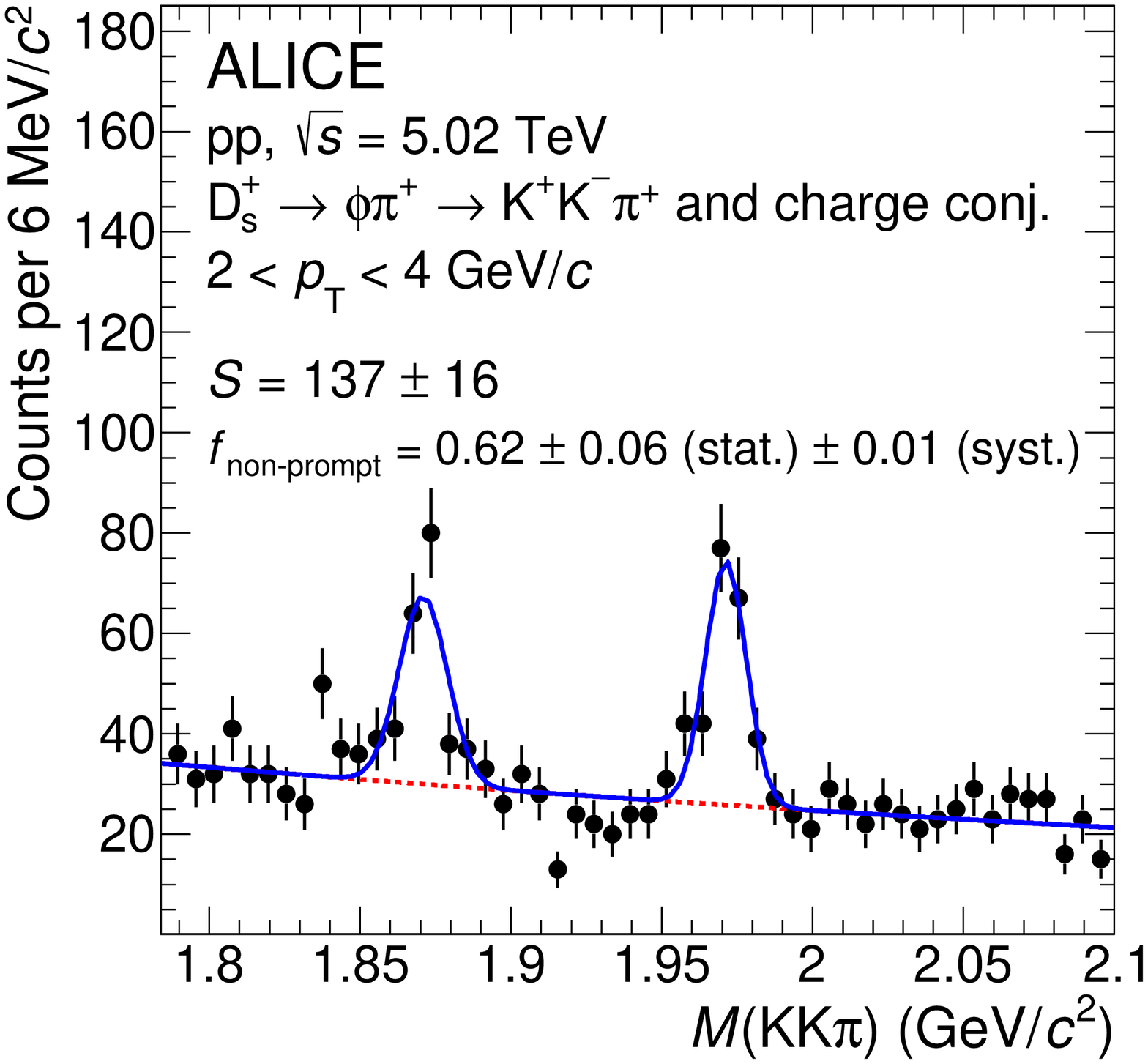}
    \end{center}
    \caption{Invariant-mass distributions of $\Dzero$, $\Dplus$, and $\Ds$ candidates and charge conjugates in $1<\pt<2~\gevc$, $8<\pt<10~\gevc$, and $2<\pt<4~\gevc$ intervals, respectively. The blue solid lines show the total fit functions as described in the text and the red dashed lines are the combinatorial background. In case of the $\Dzero$ candidates, the grey dashed line represents the combinatorial background with the contribution of the reflections. The raw-yield ($S$) values are reported together with their statistical uncertainties resulting from the fit. The fraction of non-prompt candidates in the measured raw yield is reported with its statistical and systematic uncertainties.}
    \label{fig:non_prompt_inv_mass}
\end{figure}

Samples enhanced with non-prompt candidates were selected by requiring a low candidate probability to be combinatorial background and a high probability to be non-prompt. The raw yields of $\Dzero$, $\Dplus$, and $\Ds$ mesons, including both particles and antiparticles, were extracted from binned maximum-likelihood fits to the invariant-mass ($M$) distributions. The raw yields could be extracted in transverse-momentum intervals in the range $1<\pt<24~\gev/c$ for $\Dzero$ mesons, $2<\pt<16~\gev/c$ for $\Dplus$ mesons, and $2<\pt<12~\gev/c$ for $\Ds$ mesons. 
The fit function was composed of a Gaussian for the description of the signal and of an exponential term for the background. To improve the stability of the fits, the widths of the D-meson signal peaks were fixed to the values extracted from data samples dominated by prompt candidates, given the naturally larger abundance of prompt compared to non-prompt D mesons. For the $M({\rm KK\pi})$ distribution, an additional Gaussian was used to describe the peak due to the decay $\DplustoKKpi$, with a branching ratio of ($9.68\pm 0.18)\times 10^{-3}$~\cite{Zyla:2020zbs}, present at a lower invariant-mass value than the $\Ds$-meson signal peak. For the $\Dzero$ meson, the contribution of signal candidates to the invariant-mass distribution with the wrong mass assigned to the $\Dzero$-decay tracks (reflections) was included in the fit. It was estimated based on the invariant-mass distributions of the reflected signal in the simulation, which were described as the sum of two Gaussian functions. The contribution of reflections to the raw yield is about $0.5\%-4\%$, depending on $\pt$.
Examples of invariant-mass distributions together with the result of the fits and the estimated non-prompt fractions are reported in Fig.~\ref{fig:non_prompt_inv_mass}, for the $1<\pt<2~\gevc$, $8<\pt<10~\gevc$, and $2<\pt<4~\gevc$ intervals of the $\Dzero$, $\Dplus$, and $\Ds$ candidates, respectively. The procedure used to calculate the fraction of non-prompt candidates present in the extracted raw yields is described in Section~\ref{sec:non_prompt_estimation}. The measured raw yields, although dominated by non-prompt candidates, still contain a residual contribution of prompt D mesons which satisfy the BDT-based selections. The statistical significance of the observed signals, $S/\sqrt{S+B}$, varies from 4 to 10, depending on the D-meson species and on the $\pt$ interval. 

The $\pt$-differential cross section of non-prompt D mesons was computed for each $\pt$ interval as
\begin{equation}
  \label{eq:dNdpt}
  \frac{{\rm d}^2 \sigma^{\rm D}}{{\rm d}\pt {\rm d} y}=
  \frac{1}{c_{\Delta y}(\pt)\Delta \pt}\times\frac{1}{{\rm BR}}  \times\frac{\frac{1}{2} \, \left.\fnonprompt(\pt)\times N^{\rm D+\overline D,raw}(\pt)\right|_{|y|<y_{\rm fid}(\pt)}}{ \effNP{}(\pt)} \frac{1}{\lumi} \,.
\end{equation}
The raw-yield values (sum of particles and antiparticles, $N^{\rm D+\overline D,raw}$) were divided by a factor of two and multiplied by the non-prompt fraction $\fnonprompt$ to obtain the charged-averaged yields of non-prompt D mesons. Furthermore, they were divided by the acceptance times efficiency of non-prompt D mesons
$\effNP{}$, the BR of the decay channel, 
the width of the $\pt$ interval ($\Delta \pt$), the correction factor for the rapidity coverage $c_{\Delta y}$ (see below), and the integrated luminosity $\lumi=N_{\rm ev} / \sigma_{\rm MB} $, where $N_{\rm ev}$ is the number of analysed events and $\sigma_{\rm MB}=(50.9\pm 0.9)$~mb is the cross section for the MB trigger condition~\cite{ALICE-PUBLIC-2018-014}.

\begin{figure}[tb]
    \begin{center}
    \includegraphics[width = \textwidth]{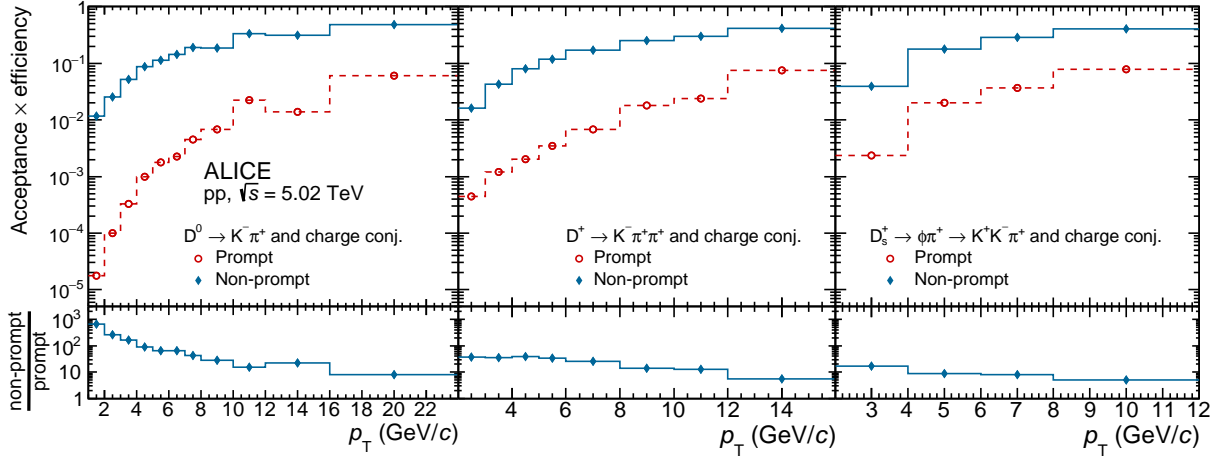}
    \end{center}
    \caption{Acceptance-times-efficiency factor for $\Dzero$, $\Dplus$, and $\Ds$ mesons as a function of $\pt$. The $(\rm Acc \times \epsilon)$ factors for non-prompt (blue) and prompt (red) D mesons are shown together with their ratio (bottom panels).}
    \label{fig:non_prompt_eff}
\end{figure}

The $(\rm Acc \times \epsilon)$ correction was obtained from simulations, described in Section~\ref{sec:apparatus}, using samples not employed in the BDT training.
The $(\rm Acc \times \epsilon)$ factors, computed for the selections used in the final result, as a function of $\pt$ for prompt and non-prompt $\Dzero$, $\Dplus$, and $\Ds$ mesons within the fiducial acceptance region are shown in Fig.~\ref{fig:non_prompt_eff}, along with the ratios of the non-prompt over prompt factors. The selection applied to obtain the non-prompt enhanced samples strongly suppresses the prompt D-meson efficiency, while the acceptance is the same between prompt and non-prompt D mesons. The prompt D-meson acceptance times efficiency is smaller than the one of non-prompt D mesons by a factor varying from 5 to 700, depending on the D-meson species and the $\pt$ interval. The difference between the $(\rm Acc \times \epsilon)$ factors of prompt and non-prompt mesons is less pronounced for $\Dplus$ than for $\Dzero$, due to the more similar lifetimes of $\Dplus$ and beauty hadrons. For $\Ds$ mesons, looser selections than those used for the other D-meson species were applied due to the lower yield of $\Ds$ mesons, leading to a smaller difference between the $(\rm Acc \times \epsilon)$ factors of the prompt and non-prompt components.

The correction factor for the rapidity acceptance $c_{\Delta y}$ was computed with FONLL perturbative QCD calculations, which have shown a good description of the rapidity dependence of the D-meson cross section~\cite{Aaij:2016jht,Acharya:2019mgn}. The correction factor was defined as the ratio between the generated D-meson yield in $\Delta y = 2\,y_{\rm fid}$ and that in $|y|<0.5$. Calculations of $c_{\Delta y}$  based on the PYTHIA~8 event generator were in agreement within 1\%.
The $\fnonprompt$ fraction was calculated with a novel data-driven approach, which is described in Section~\ref{sec:non_prompt_estimation}.

\subsection{Data-driven estimation of non-prompt fraction}
\label{sec:non_prompt_estimation}
The fraction $\fnonprompt$ of non-prompt D mesons in the raw yield was estimated by sampling the raw yield at different values of the BDT output related to the candidate probability of being a non-prompt D meson. In this way, a set of raw yields $\rawY{_\mathrm{i}}$ with different contributions of prompt and non-prompt D mesons was obtained. These raw yields can be related to the corrected yields of prompt ($\Np$) and non-prompt ($\Nnp$) D mesons via the acceptance-times-efficiency factors as follows
%
%
\begin{equation}
   \effP{\rm i}\times \Np +  \effNP{\rm i}\times \Nnp - \rawY{\rm i} = \delta_\mathrm{i}.
\label{eq:eq_set}
\end{equation}
In the above equation, $\delta_\mathrm{i}$ represents a residuum that accounts for the equation not holding exactly due to the uncertainty on $\rawY{\rm i}$, $\effNP{\rm i}$, and $\effP{\rm i}$. The definition of $n$ selections leads to the following algebraic system
\begin{equation}
\renewcommand\arraystretch{1.3} {
\left(
\begin{array}{cc}
\effP{1} & \effNP{1}\\
\vdots & \vdots\\
\effP{n} & \effNP{n}
\end{array}
 \right)} 
 \times
 \left(
\begin{array}{c}
 \Np\\
 \Nnp
\end{array}
 \right) 
 -
  \left(
\begin{array}{c}
\rawY{1}\\
\vdots\\
Y_n
\end{array}
 \right)
 = 
\left(
\begin{array}{c}
\delta_1\\
\vdots\\
\delta_n
\end{array}
 \right)
 ,
\label{eq:systemCutVar}
\end{equation}
that can be exactly solved in case of two equations (assuming $\delta_i=0$). 
With $n$ selections, the $\Np$ and $\Nnp$ parameters are obtained by minimising the $\chi^2$ \begin{equation}
\label{eq:chisq}
 \chi^2 = \pmb{\delta^T}\pmb{C^{-1}}\pmb{\delta},
\end{equation}
where $\pmb{\delta^T}$ is the row vector of residuals and $\pmb{C}$ the covariance matrix
accounting for the uncertainties inherent to each equation. The variances $\sigma_\mathrm{i}^2$ were calculated from the statistical uncertainty on the raw yields and efficiencies as
\begin{equation}
 \sigma_\mathrm{i}^2 = \sigma_{\rawY{\rm i}}^2 + \Np\times\sigma^2_{\effP{\rm i}} + \Nnp\times\sigma^2_{\effNP{\rm i}}.
\end{equation}
Given that the corrected yields are unknown variables, an iterative procedure was used to define the total uncertainty: in the first step 
only the uncertainty on the raw yields was taken into account, while from the second iteration 
the corrected yields $\Np$ and $\Nnp$ obtained in the previous step were also used. In the covariance terms $\sigma_\mathrm{i, j}$ 
the correlation coefficient was assumed to be
\begin{equation}
    \rho_{\rm i, j} = \frac{\sigma_\mathrm{i}}{\sigma_{\rm j}}\text{,~with~}{\rm i}\subset {\rm j}.
\label{eq:correlation}
\end{equation}
This assumption is justified by the fact that the BDT response is sampled monotonically, so that the $n$ selections are ordered in such a way that the $\mathrm{i^{th}}$ selected sample is completely included in the $\mathrm{(i-1)^{th}}$ one. 
For $\Dzero$ mesons, only the equation for the strictest set of selections was defined as in Eq.~\ref{eq:eq_set}. All the others were expressed in terms of the difference between the $\mathrm{(i-1)^{th}}$ and the $\mathrm{i^{th}}$ raw yields, $\Delta\rawY{\rm i-1,i} = \rawY{\rm i-1}-\rawY{\rm i}$. In this case, the covariance terms were assumed to be zero, resulting in a diagonal covariance matrix.

The fraction of non-prompt D mesons in the raw yield can be computed for any set of selections i from the corrected yields obtained from the $\chi^2$ minimisation as
\begin{equation}
\label{eq:fnpromptSystem}
  \fnonprompt^{\rm i} = \frac{\effNP{\rm i} \times \Nnp}{\effNP{\rm i} \times \Nnp+\effP{\rm i} \times \Np}.
\end{equation}
Rather than from the $\Nnp$ parameter obtained from the minimisation of the $\chi^{2}$ in Eq.~\ref{eq:chisq}, the final values of the non-prompt D-meson cross sections were determined by choosing a selection providing a high non-prompt component and a good signal extraction, as described in Section~\ref{sec:analysis}, and by calculating its respective $\fnonprompt$ fraction according to Eq.~\ref{eq:fnpromptSystem}. This approach facilitates the determination of the systematic uncertainty.

\begin{figure}[tb]
    \begin{center}
    \offinterlineskip
    \includegraphics[width = 0.45\textwidth]{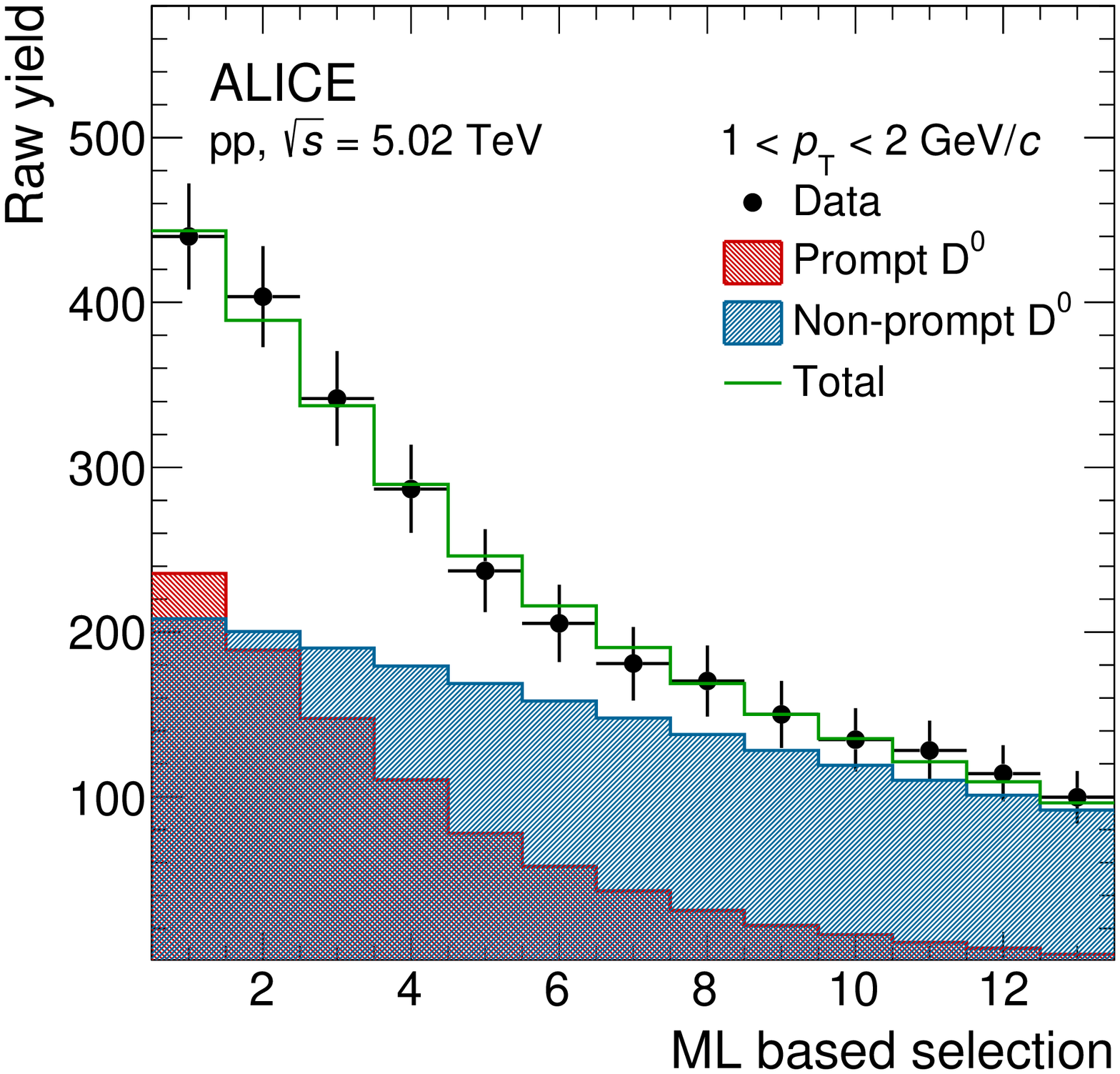}
    \includegraphics[width = 0.45\textwidth]{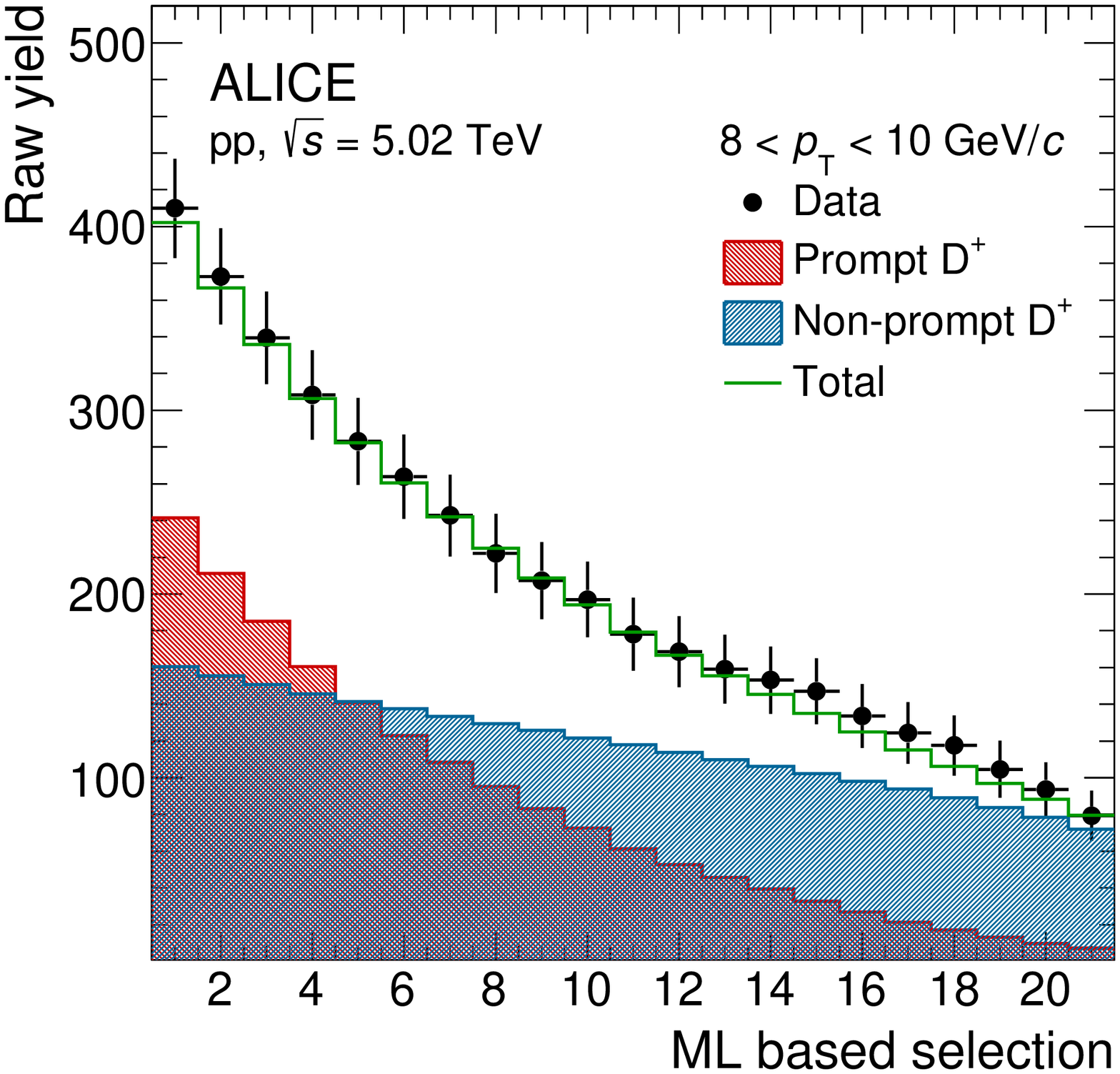}
    \includegraphics[width = 0.45\textwidth]{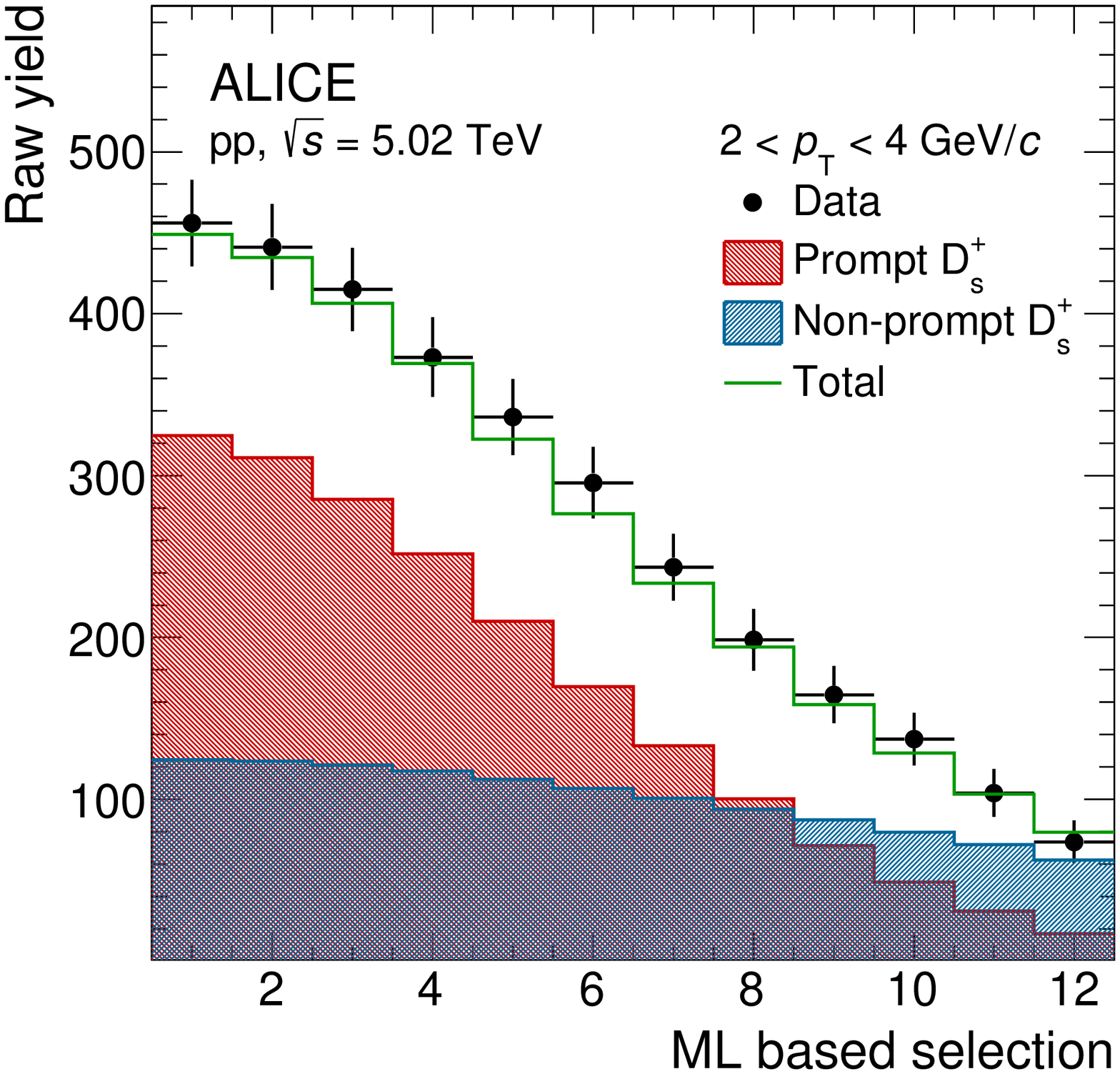}
    \includegraphics[width = 0.45\textwidth]{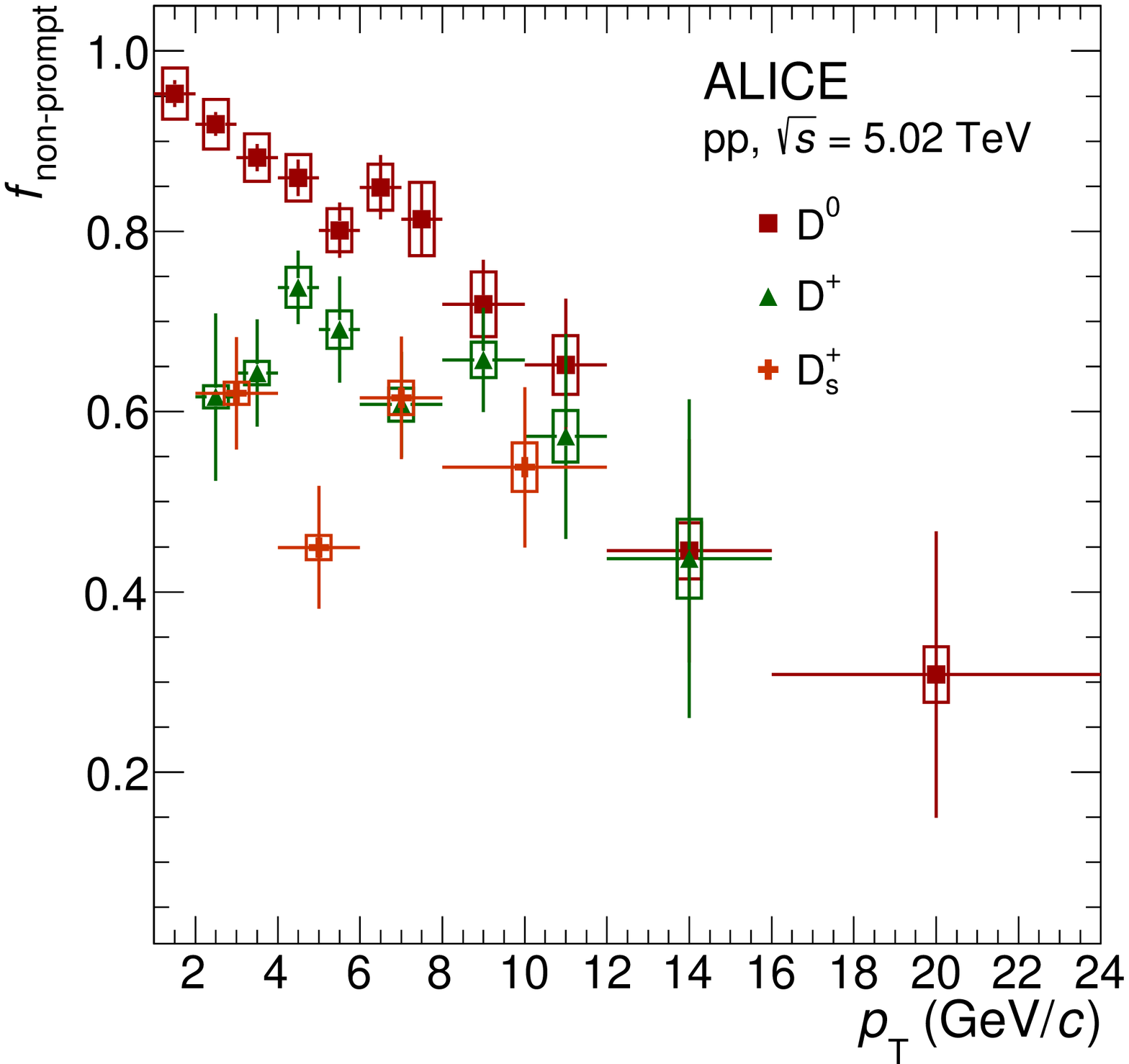}
    \end{center}
    \caption{Examples of raw-yield distribution as a function of the BDT-based selection employed in the $\chi^2$-minimisation procedure adopted for the determination of $\fnonprompt$ of $\Dzero$ mesons (top left panel), $\Dplus$ mesons (top right panel), and $\Ds$ mesons (bottom left panel) for three different $\pt$ intervals. Bottom right panel, $\fnonprompt$ fraction as a function of $\pt$ obtained for the set of selection criteria adopted in the analysis of non-prompt D mesons.}
    \label{fig:non_prompt_frac}
\end{figure}

Figure~\ref{fig:non_prompt_frac} shows an example of raw-yield distribution as a function of the BDT-based selection employed in the minimisation procedure for $\Dzero$ mesons with $1<\pt<2~\GeV/c$ (top left panel), $\Dplus$ mesons with $8<\pt<10~\GeV/c$ (top right panel), and $\Ds$ mesons with $2<\pt<4~\GeV/c$ (bottom left panel). The leftmost data point of each distribution is the raw yield corresponding to the looser selection on the BDT output related to the candidate's probability of being a non-prompt D meson, while the rightmost one corresponds to the strictest selection, which is expected to preferentially select non-prompt D mesons. The prompt and non-prompt components, obtained for each BDT-based selection from the minimisation procedure as $\effP{\rm i}\Np$ and $\effNP{\rm i}\Nnp$, are represented by the red and blue filled histograms, respectively, while their sum is reported by the green histograms. In the bottom right panel of Fig.~\ref{fig:non_prompt_frac}, the $\fnonprompt$ fractions of $\Dzero$, $\Dplus$, and $\Ds$ mesons, computed with the formula in Eq.~\ref{eq:fnpromptSystem}, corresponding to the samples enhanced with non-prompt candidates introduced in Section~\ref{sec:analysis} are shown as a function of $\pt$. The vertical bars represent the statistical uncertainty computed propagating the uncertainties on the corrected yields obtained with the $\chi^2$-minimisation procedure, where the correlation between $\Np$ and $\Nnp$ is also accounted for. The open boxes represent the systematic uncertainty, which will be described in Section~\ref{sec:systematic}.
The $\fnonprompt$ fractions range in the interval $0.3-0.95$ for $\Dzero$ mesons, $0.4-0.75$ for $\Dplus$ mesons, and $0.4-0.65$ for $\Ds$ mesons. In general, the $\fnonprompt$ values decrease with $\pt$, because at high $\pt$ a less stringent selection on the BDT probability of being non-prompt is needed to preserve a sufficient number of candidates to perform the invariant-mass analysis. 

\subsection{Measurement of prompt $\Dplus$ and $\Ds$ mesons} 
\label{sec:analysis_prompt}

The measurement of prompt $\Dplus$ and $\Ds$ mesons follows the same procedure described in Section~\ref{sec:analysis_non_prompt}. The same machine-learning models trained for the non-prompt $\Dplus$ and $\Ds$ analysis were employed. Samples containing a small fraction of non-prompt candidates were obtained selecting on the BDT outputs and requiring a low candidate probability to be combinatorial background and non-prompt. The raw yields of $\Dplus$ and $\Ds$ mesons were extracted in the transverse-momentum intervals $0<\pt<36~\gevc$ and $1<\pt<24~\gevc$, respectively, extending the measurement to lower $\pt$ with respect to the previously published results~\cite{Acharya:2019mgn}. The employed fit configurations were the same as for the non-prompt analysis, except that the widths of the $\Dplus$- and $\Ds$-meson signal peaks were unconstrained in the fit. Moreover, for $\Dplus$ mesons in $0<\pt<1~\gevc$ a third-order polynomial function was used to describe the combinatorial background, instead of an exponential function.
\begin{figure}[tb]
    \begin{center}
    \includegraphics[width = 0.45\textwidth]{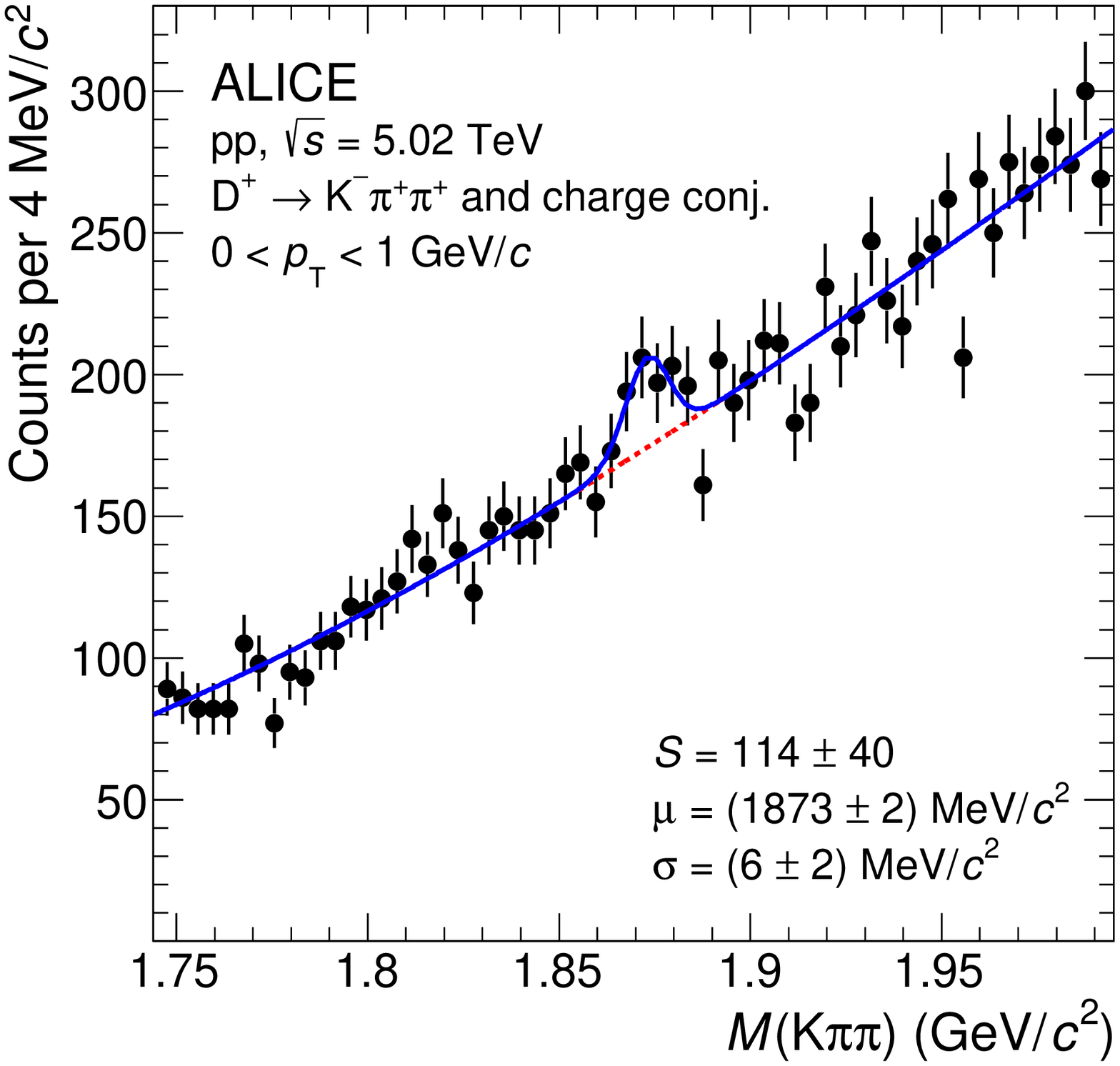}
    \includegraphics[width = 0.45\textwidth]{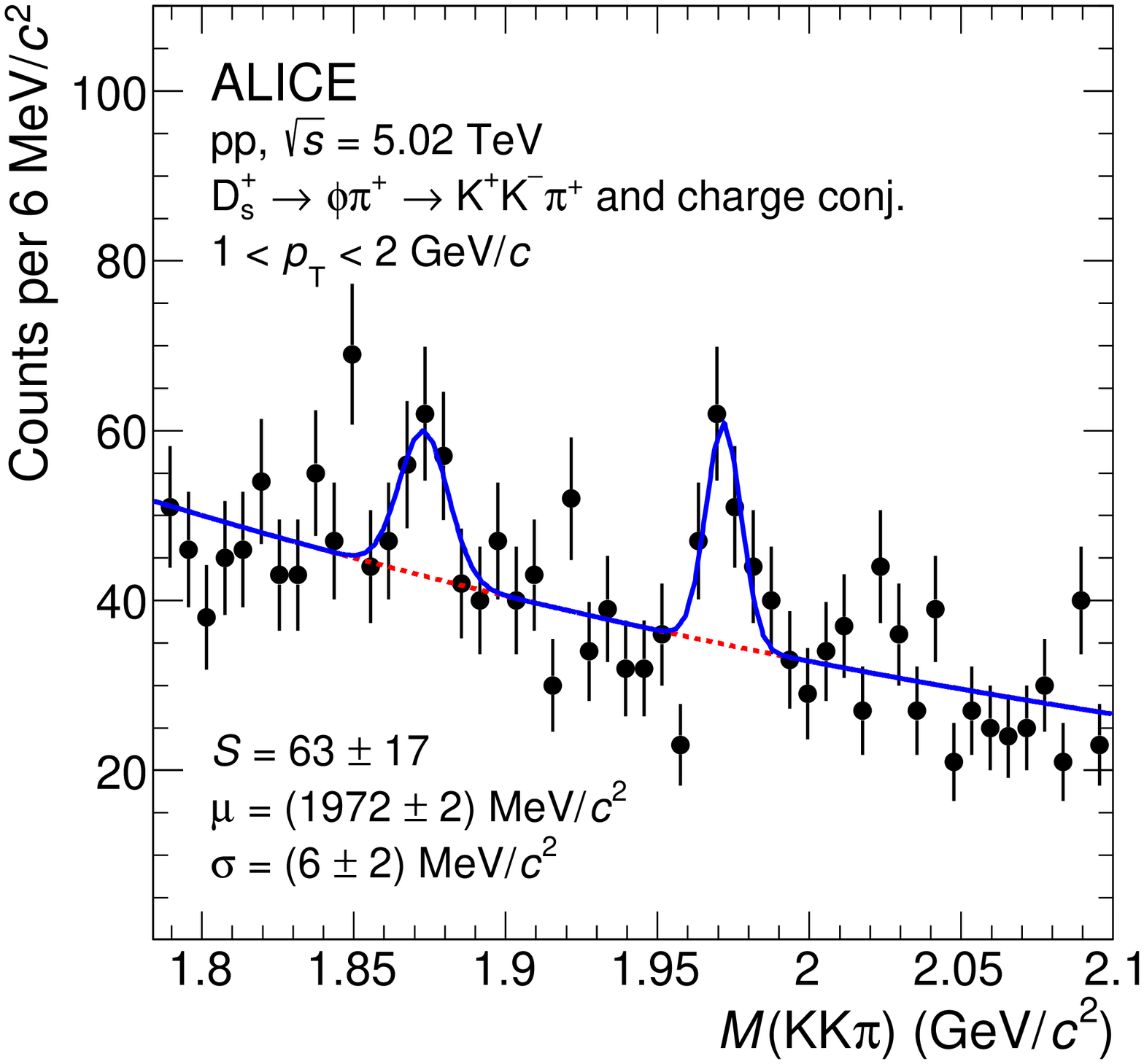}
    \end{center}
    \caption{Invariant-mass distributions of $\Dplus$ and $\Ds$ candidates and charge conjugates in the intervals $0<\pt<1~\gevc$ and $1<\pt<2~\gevc$, respectively. The blue solid lines show the total fit functions as described in the text and the red dashed lines are the combinatorial-background components. The values of the mean ($\mu$) and the width ($\sigma$) of the signal peak are reported together with the raw yield ($S$). The reported uncertainties are only the statistical uncertainties from the fit.}
    \label{fig:prompt_inv_mass}
\end{figure}
Figure~\ref{fig:prompt_inv_mass} shows the invariant-mass distributions, together with the result of the fits, in the $0<\pt<1~\gevc$ and $1<\pt<2~\gevc$ intervals for $\Dplus$ and $\Ds$ candidates, respectively. The statistical significance of the observed signals varies from about 3 to 40 for $\Dplus$ mesons and from 4 to 14 for $\Ds$ mesons, depending on the $\pt$ interval. The $S/B$ values obtained are $0.07-2.5$ ($0.31-3.1$) for $\Dplus$ ($\Ds$) mesons, depending on $\pt$. The performance of the adopted BDT-based selections was compared with that obtained in the previous study~\cite{Acharya:2019mgn}. An improvement of the statistical significance by a factor $1.1-2$ ($1.2-1.7$) for $\Dplus$ ($\Ds$) mesons in the common $\pt$ regions of the two measurements is observed, implying a reduction of statistical uncertainties by the same factor. Furthermore, the efficiency for prompt $\Dplus$ and $\Ds$ mesons is higher in the BDT-based analysis by a factor $1.2-4$ and $1.7-2.2$, respectively, depending on the $\pt$ interval.

\begin{figure}[tb]
    \begin{center}
    \includegraphics[width = 0.45\textwidth]{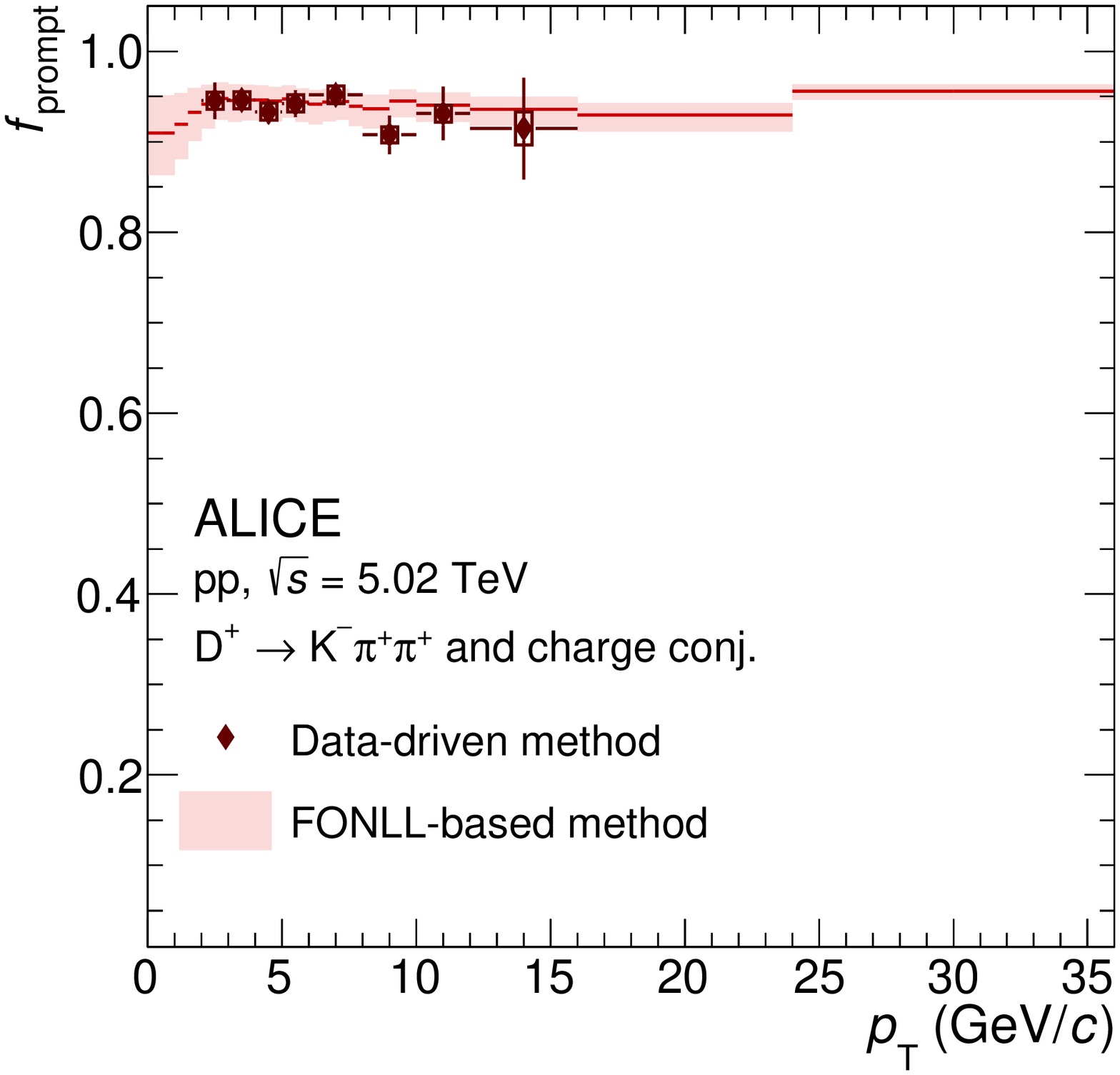}
    \includegraphics[width = 0.45\textwidth]{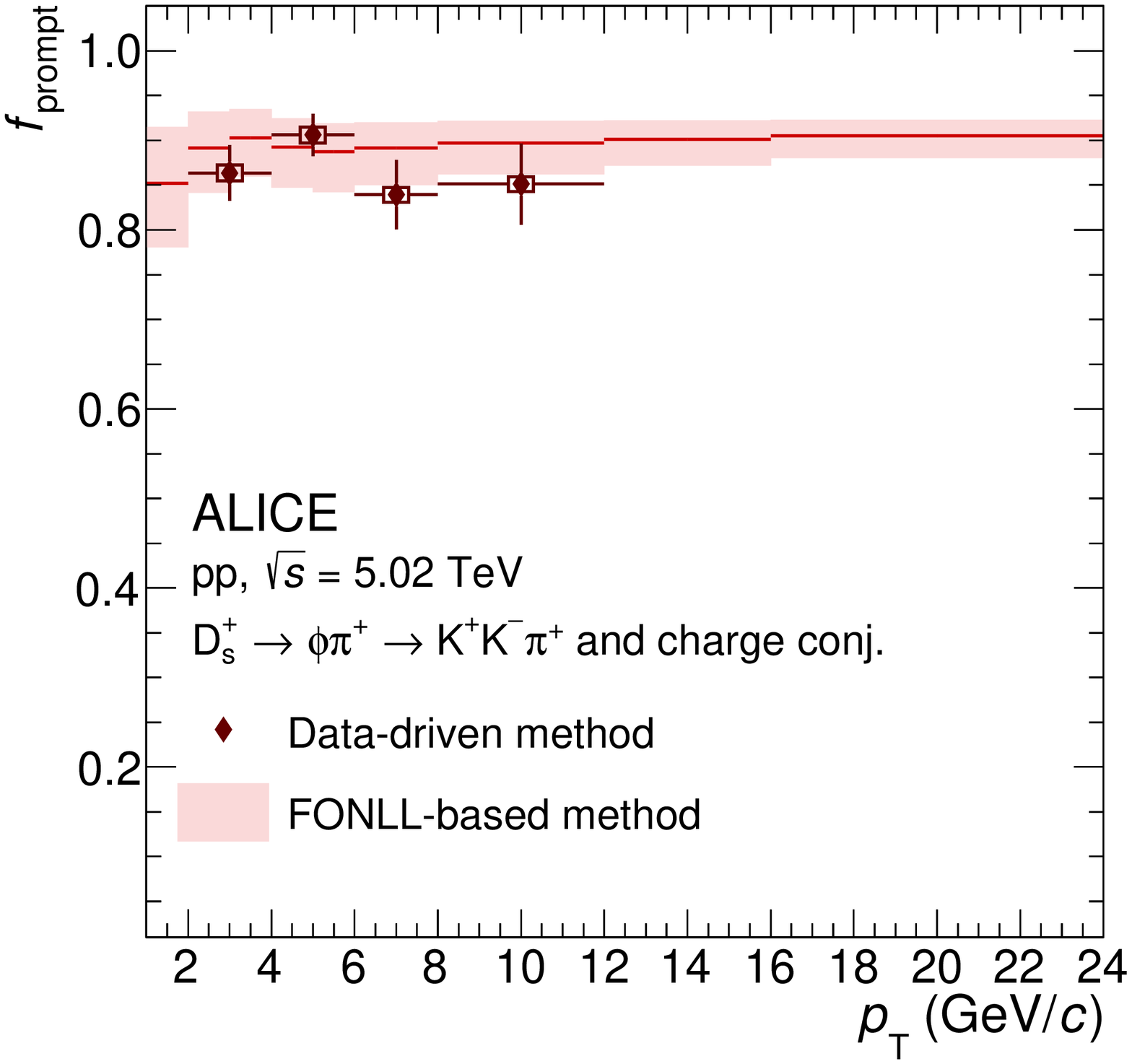}
    \end{center}
    \caption{Comparison of the fractions of prompt $\Dplus$- and $\Ds$-meson raw yields as a function of $\pt$ between the FONLL-based and the data-driven approach. The results from the data-driven method are shown as diamond markers with the error bars (boxes) representing the statistical (systematic) uncertainty. The central values of $\fprompt$ from the FONLL-based approach are shown by the continuous line and their uncertainty by the shaded boxes.}
    \label{fig:prompt_frac_comp}
\end{figure}

The data-driven method described in Section~\ref{sec:non_prompt_estimation}, which is based on the reliable extraction of raw yields with different fractions of prompt and non-prompt candidates, cannot be used for the estimation of the $\fprompt$ fraction in all the $\pt$ intervals of the prompt $\Dplus$ and $\Ds$ measurements, due to the limited size of the analysed data sample.
Thus, the $\fprompt$ fraction was calculated similarly to previous measurements (see e.g. Refs.~\cite{Acharya:2017jgo,Acharya:2018hre}) using the beauty-hadron production cross sections from FONLL calculations, the beauty~hadron to $\mathrm{D+X}$ decay kinematics from the PYTHIA~8 decayer, and the acceptance-time-efficiency correction factors for non-prompt $\Dplus$ and $\Ds$ mesons from Monte Carlo simulations. The values of $\fprompt$ range between 0.86 and 0.96 depending on the D-meson species and $\pt$ interval. The procedure to estimate the systematic uncertainty on $\fprompt$ will be described in Section~\ref{sec:systematic}. Figure~\ref{fig:prompt_frac_comp} reports the $\Dplus$- and $\Ds$-meson $\fprompt$ fractions obtained with the FONLL-based approach compared with those resulting from the data-driven method,  the latter were computed in the $\pt$ ranges of the non-prompt $\Dplus$ and $\Ds$ measurements where a good reliability of the method can be granted. The fractions of prompt D-meson yields estimated with the two different strategies are well in agreement within the statistical and systematic uncertainties in the common $\pt$ intervals.

\section{Systematic uncertainties}
\label{sec:systematic}

The systematic uncertainties on the measurement of prompt and non-prompt D-meson cross sections were estimated with procedures similar to those described in~Refs.~\cite{Acharya:2019mgn, Acharya:2019mno,Acharya:2018hre}, including the following sources: (i) extraction of the raw yield from the invariant-mass distributions; 
(ii) non-prompt and prompt fraction estimations;
(iii) track reconstruction efficiency; 
(iv) D-meson selection efficiency; 
(v) PID efficiency; 
(vi) generated D-meson $\pt$ shape in the simulation.
In addition, an overall normalisation systematic uncertainty induced by the branching ratios of the considered D-meson decays~\cite{Zyla:2020zbs} and the integrated luminosity~\cite{ALICE-PUBLIC-2018-014} were considered. The estimated values of the systematic uncertainties for some representative $\pt$ intervals of the different analyses are summarised in Table~\ref{tab:sysunc_yieldtable}. The contributions of the different sources were summed in quadrature to obtain the total systematic uncertainty. For non-prompt D mesons, the systematic uncertainties on the non-prompt fraction estimation and the raw-yield extraction were treated as correlated and summed linearly.

\begin{table}[tb]
\caption{Summary of the relative systematic uncertainties on non-prompt $\Dzero$, $\Dplus$, and $\Ds$ cross sections and prompt $\Dplus$ and $\Ds$ cross sections in different $\pt$ intervals.}
\centering
\scalebox{0.88}{ 
\renewcommand*{\arraystretch}{1.2}
\begin{tabular}[t]{l|>{\centering}p{0.05\linewidth}>{\centering}p{0.05\linewidth}|>{\centering}p{0.05\linewidth}>{\centering}p{0.05\linewidth}|>{\centering}p{0.05\linewidth}>{\centering}p{0.05\linewidth}|>{\centering}p{0.05\linewidth}>{\centering}p{0.05\linewidth}|>{\centering}p{0.05\linewidth}>{\centering\arraybackslash}p{0.06\linewidth}}
\toprule
 Meson& \multicolumn{2}{c|}{non-prompt $\Dzero$} & \multicolumn{2}{c|}{non-prompt  $\Dplus$} & \multicolumn{2}{c|}{non-prompt $\Ds$} & \multicolumn{2}{c|}{prompt  $\Dplus$} &\multicolumn{2}{c}{prompt $\Ds$}\\
$\pt~(\GeV/c)$  & $1-2$   & \multicolumn{1}{c|}{$10-12$} & $2-3$   & \multicolumn{1}{c|}{$10-12$} & $2-4$   & \multicolumn{1}{c|}{$8-12$}  & $0-1$  & \multicolumn{1}{c|}{$10-12$} & $1-2$           & $8-12$         \\
\midrule
Signal yield                              & 5\%   & \multicolumn{1}{c|}{7\%}   & 3\%   & \multicolumn{1}{c|}{5\%}   & 4\%   & \multicolumn{1}{c|}{3\%}   & 10\% & \multicolumn{1}{c|}{3\%}   & 7\%           & 3\%           \\
\multicolumn{1}{l|}{Tracking efficiency}  & 3\%  & \multicolumn{1}{c|}{5\%}   & 5\%   & \multicolumn{1}{c|}{7\%}   & 5\%   & \multicolumn{1}{c|}{7\%}   & 4\%  & \multicolumn{1}{c|}{7\%}   & 4\%           & 7\%           \\
\multicolumn{1}{l|}{Selection efficiency} & 10\%   & \multicolumn{1}{c|}{5\%}   & 10\%  & \multicolumn{1}{c|}{5\%}   & 7\%   & \multicolumn{1}{c|}{5\%}   & 10\% & \multicolumn{1}{c|}{2\%}   & 8\%           & 3\%           \\
PID efficiency                            & 0     & \multicolumn{1}{c|}{0}     & 0     & \multicolumn{1}{c|}{0}     & 0     & \multicolumn{1}{c|}{0}     & 0    & \multicolumn{1}{c|}{0}     & 0             & 0             \\
$\pt$ shape in MC                            & 1\%   & \multicolumn{1}{c|}{0}     & 1\%   & \multicolumn{1}{c|}{1\%}   & 1\%   & \multicolumn{1}{c|}{1\%}   & 7\%  & \multicolumn{1}{c|}{0}     & 1\%           & 0           \\
Fraction estimation                       & 3\%   & \multicolumn{1}{c|}{5\%}   & 2\%   & \multicolumn{1}{c|}{5\%}   & 2\%   & \multicolumn{1}{c|}{4\%}   & $^{+4}_{-4}$\%  & \multicolumn{1}{c|}{$^{+1}_{-2}$\%}   & $^{+6}_{-7}$\%           & $^{+2}_{-3}$\%          \\
Branching ratio                           & \multicolumn{2}{c|}{1\%}           & \multicolumn{2}{c|}{2\%}           & \multicolumn{2}{c|}{4\%}           & \multicolumn{2}{c|}{2\%}          & \multicolumn{2}{c}{4\%}       \\
\midrule
Luminosity                                & \multicolumn{10}{c}{2\%}\\
\bottomrule
\end{tabular}
}
\label{tab:sysunc_yieldtable}	
\end{table}

The systematic uncertainty on the raw-yield extraction was evaluated by repeating the fit of the invariant-mass distribution varying the lower and upper limits of the fit range and the functional form of the background fit function. In order to test the sensitivity to the line-shape of the signal, a bin-counting method was used, in which the signal yield was obtained by integrating the invariant-mass distribution after subtracting the background estimated from the side-band fit. In addition, for the analysis of non-prompt D mesons the width of the Gaussian function used to model the signal peaks was varied within the uncertainty of the value obtained from the fits to the prompt-enhanced sample. The effect was found to be negligible, hence no additional systematic uncertainty was assigned. For non-prompt $\Dzero$ mesons, an additional contribution due to the description of signal reflections in the invariant-mass distribution was estimated by varying the shape and the normalisation of the templates used for the reflections in the invariant-mass fits. The systematic uncertainty was defined as the RMS of the distribution of the signal yields obtained from all these variations and ranges from 1\% to 11\% depending on the D-meson species and the $\pt$ interval.

The systematic uncertainty on the value of $\fnonprompt$ obtained with the data-driven approach was estimated by changing the sets of selection criteria used for the procedure described in Section~\ref{sec:non_prompt_estimation}. A systematic uncertainty ranging from 2\% to 10\% was assigned. This source of systematic uncertainty was found to be mostly correlated with the signal extraction procedure. The correlation was evaluated by repeating the computation of $\fnonprompt$ varying the fit configurations used for the raw-yield extraction, as described above. 
For the analysis of prompt $\Dplus$ and $\Ds$ mesons, the systematic uncertainty on $\fprompt$ was estimated by varying the FONLL parameters (b-quark mass, factorisation, and renormalisation scales) as prescribed in~\cite{Cacciari:2012ny}. It ranges between $^{+1}_{-1}$\% and $^{+6}_{-7}$\% depending on the D-meson species and $\pt$ interval.

The systematic uncertainty on the track reconstruction efficiency was evaluated by varying the track-quality selection criteria and by comparing the prolongation probability of the TPC tracks to the ITS hits in data and simulation. The comparison of the ITS-TPC prolongation efficiency in data and simulations was performed after weighting the relative abundances of primary and secondary particles in the simulation to match those observed in data, which were estimated via fits to the inclusive track impact-parameter distributions~\cite{ALICE-PUBLIC-2017-005}. The estimated uncertainty depends on the D-meson $\pt$ and ranges from 3\% to 5\% for the two-body decay of $\Dzero$ mesons and from 4\% to 7\% for the three-body decays of $\Dplus$ and $\Ds$ mesons. 

The systematic uncertainty on the selection efficiency originates from imperfections in the description of the detector resolutions and alignments in the simulation. It was estimated by comparing the corrected yields obtained by repeating the analysis with different machine-learning selection criteria, i.e. varying the selections on the BDT outputs, resulting in a significant modification of the efficiencies, raw yield and background values. The assigned systematic uncertainty ranges from 2\% to 10\%.

To estimate the uncertainty on the PID selection efficiency, the pion and kaon PID selection efficiencies were compared in data and in simulations. For this study, a pure sample of pions was selected from $\kzero$ and $\lmb$ decays, while samples of kaons in the TPC (TOF) were obtained applying a strict PID selection using the TOF (TPC) information. Since no significant differences were observed, no systematic uncertainty was assigned. As an additional test, the analysis was repeated without PID selection. The resulting D-meson cross sections were found to be compatible with those obtained with the PID selection.

The systematic effect on the efficiency due to a possible difference between the real and simulated D-meson transverse-momentum distributions was estimated by evaluating the efficiency after reweighting the $\pt$ shape from the PYTHIA~8 generator to match the one from FONLL calculations. The weights were applied to the $\pt$ distributions of prompt D mesons and to the decaying beauty hadrons in case of non-prompt D mesons. The assigned uncertainty is 7\%  in the $\pt$ interval $0-1~\gevc$ of the prompt $\Dplus$ meson, where the selection criteria are strict, while for other $\pt$ intervals the uncertainty is less than 1\%.

\section{Results}
\label{sec:results}
\subsection{Production cross sections}
\label{sec:NPD_cross_sec}
\begin{figure}[!tb]
    \begin{center}
    \includegraphics[width = 0.48\textwidth]{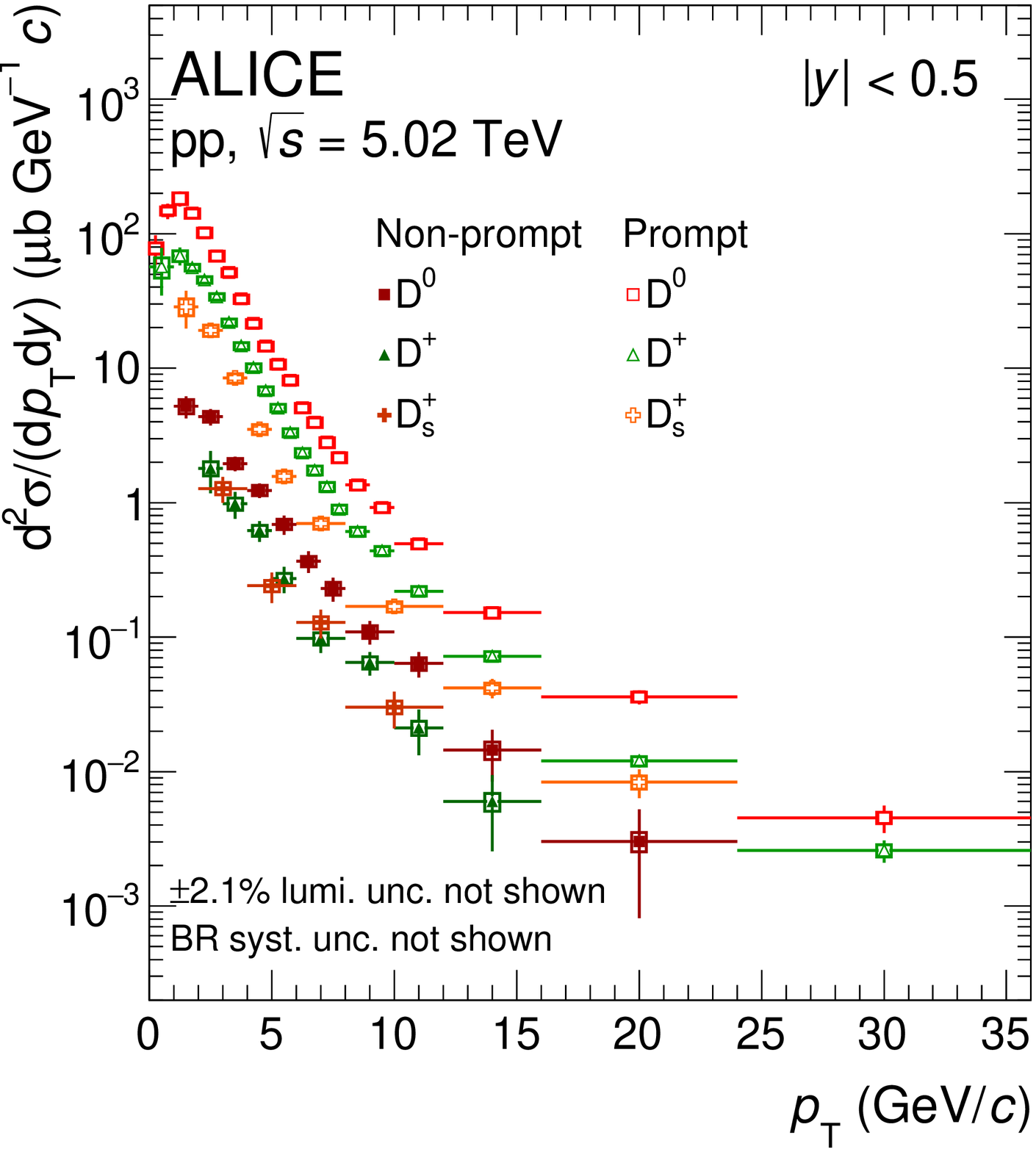}
    \includegraphics[width = 0.48\textwidth]{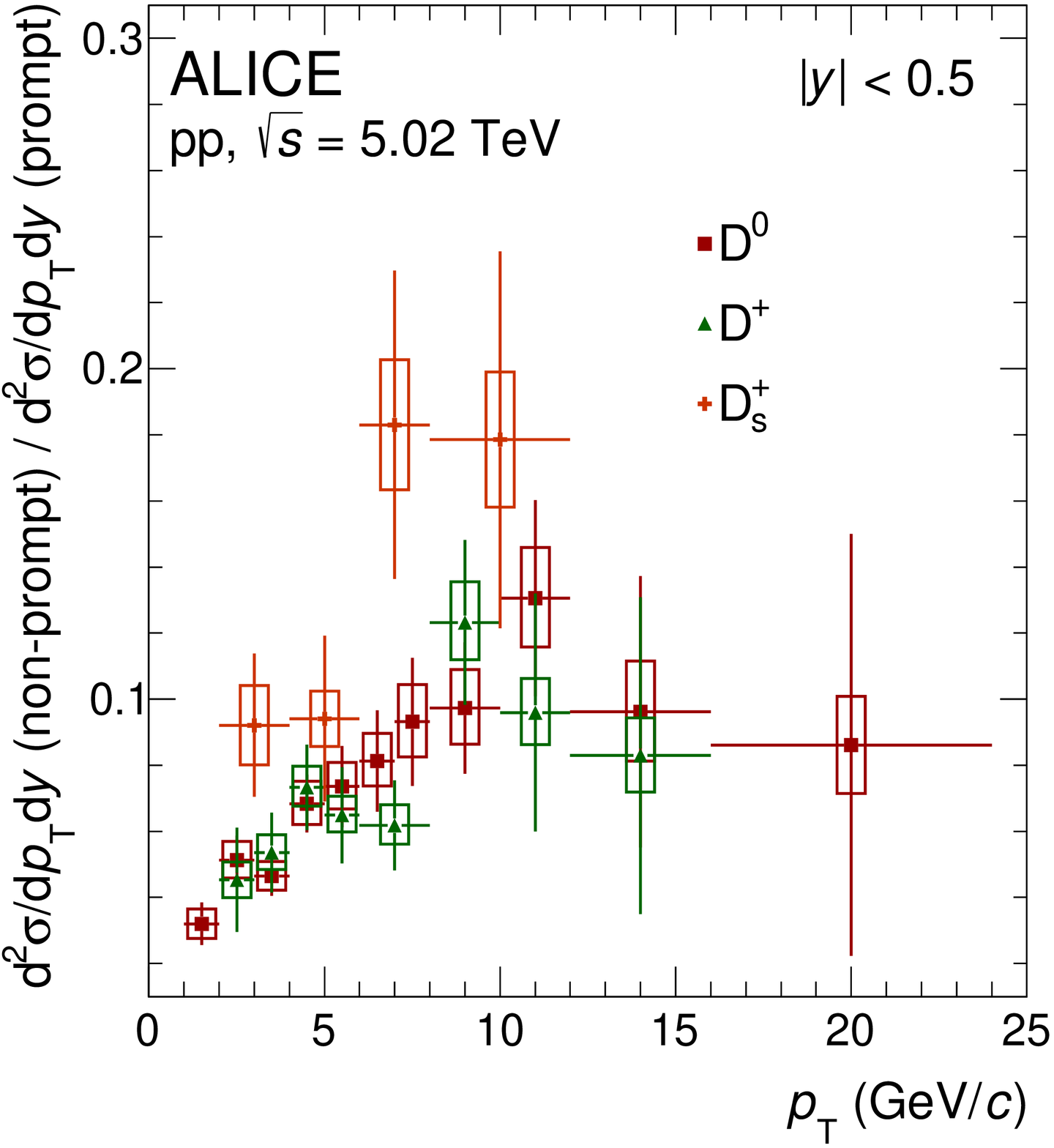}
    \end{center}
    \caption{Left: $\pt$-differential production cross sections of prompt and non-prompt $\Dzero$, $\Dplus$, and $\Ds$ mesons in pp collisions at $\s=5.02~\TeV$. The measurement of prompt $\Dzero$ mesons is the one reported in Ref.~\cite{Acharya:2019mgn}, with updated decay BR as discussed in the text. Right: ratios of $\pt$-differential production cross sections of non-prompt and prompt $\Dzero$, $\Dplus$, and $\Ds$ mesons. Statistical (vertical bars) and systematic uncertainties (boxes) are shown. The symbols are positioned horizontally at the centre of each $\pt$ interval, with the horizontal bars representing the width of the $\pt$ interval.}
    \label{fig:pt_diff_crosssec}
\end{figure}
The $\pt$-differential production cross sections of prompt and non-prompt $\Dzero$, $\Dplus$, and $\Ds$ mesons measured in $|y|<0.5$ are shown in the left panel of Fig.~\ref{fig:pt_diff_crosssec}. The $\pt$-differential cross sections of prompt $\Dplus$ and $\Ds$ mesons are compatible within uncertainties with the previous results~\cite{Acharya:2019mgn}, but have extended $\pt$ coverage and total uncertainties reduced by a factor ranging from 1.05 to 1.6 depending on $\pt$ and D-meson species
 due to the improved analysis technique described in Section~\ref{sec:analysis_prompt}. The measurement of prompt $\Dzero$ mesons is the one reported previously in Ref.~\cite{Acharya:2019mgn}, scaled for the updated $\mathrm{BR} = (3.950\pm0.031)\%$ of the $\DzerotoKpi$ decay reported in Ref.~\cite{Zyla:2020zbs}. 

The right panel of Fig.~\ref{fig:pt_diff_crosssec} shows the ratios of the $\pt$-differential cross sections of non-prompt and prompt D mesons. The statistical uncertainties assigned to each ratio were computed considering that those of the prompt and non-prompt measurements are uncorrelated. This assumption is valid since the fraction of D-meson candidates shared by the two samples is small. The systematic uncertainty related to the determination of the tracking efficiency and to the luminosity were propagated as correlated in the ratios, while all the other sources of systematic uncertainties were considered as uncorrelated between the measurements of prompt and non-prompt D mesons. The ratio increases with increasing $\pt$ for all the three D-meson species up to $\pt=12~\GeV/c$, as expected due to the harder $\pt$ distribution of beauty hadrons ($\bhad$) compared to D mesons. The ratios for $\Dplus$ and $\Dzero$ mesons are compatible within uncertainties, while for the $\Ds$ meson the central points are systematically higher compared to the other two D-meson species, suggesting a larger contribution of beauty-hadron decays to $\Ds$ compared to non-strange D mesons, although no firm conclusion can be drawn given the current uncertainties.

\begin{figure}[!tb]
    \begin{center}
    \includegraphics[width = 0.48\textwidth]{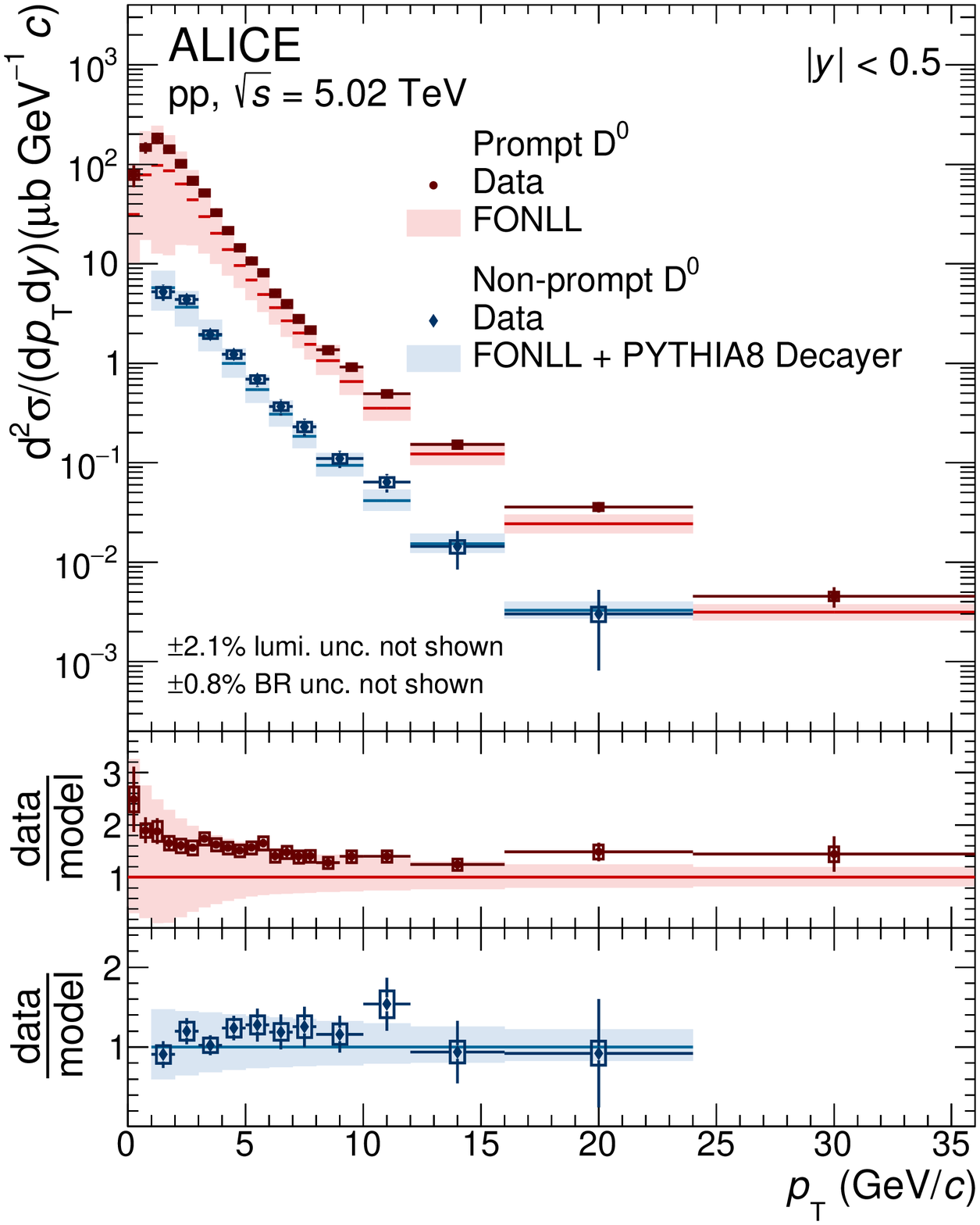}
    \includegraphics[width = 0.48\textwidth]{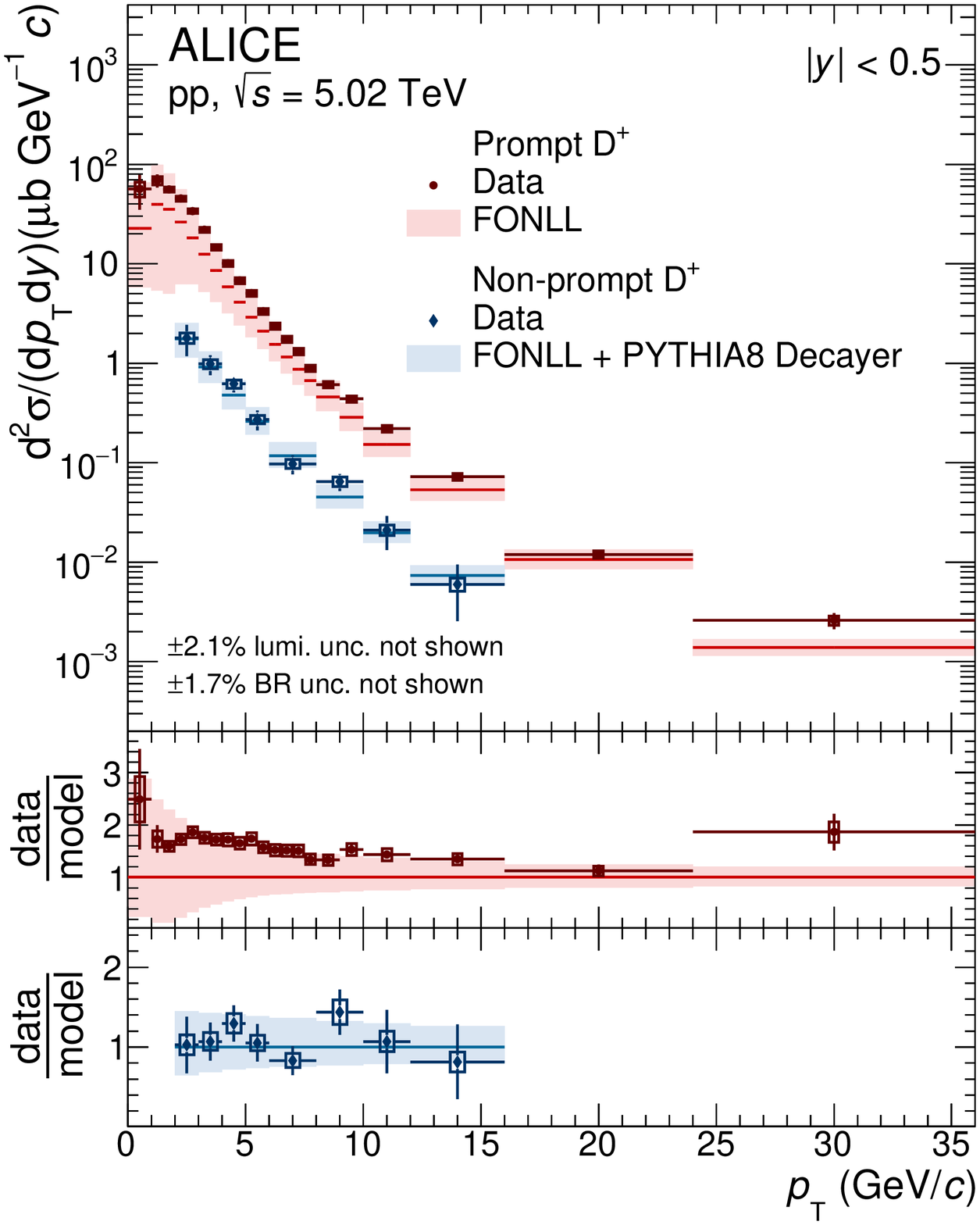}
    \includegraphics[width = 0.48\textwidth]{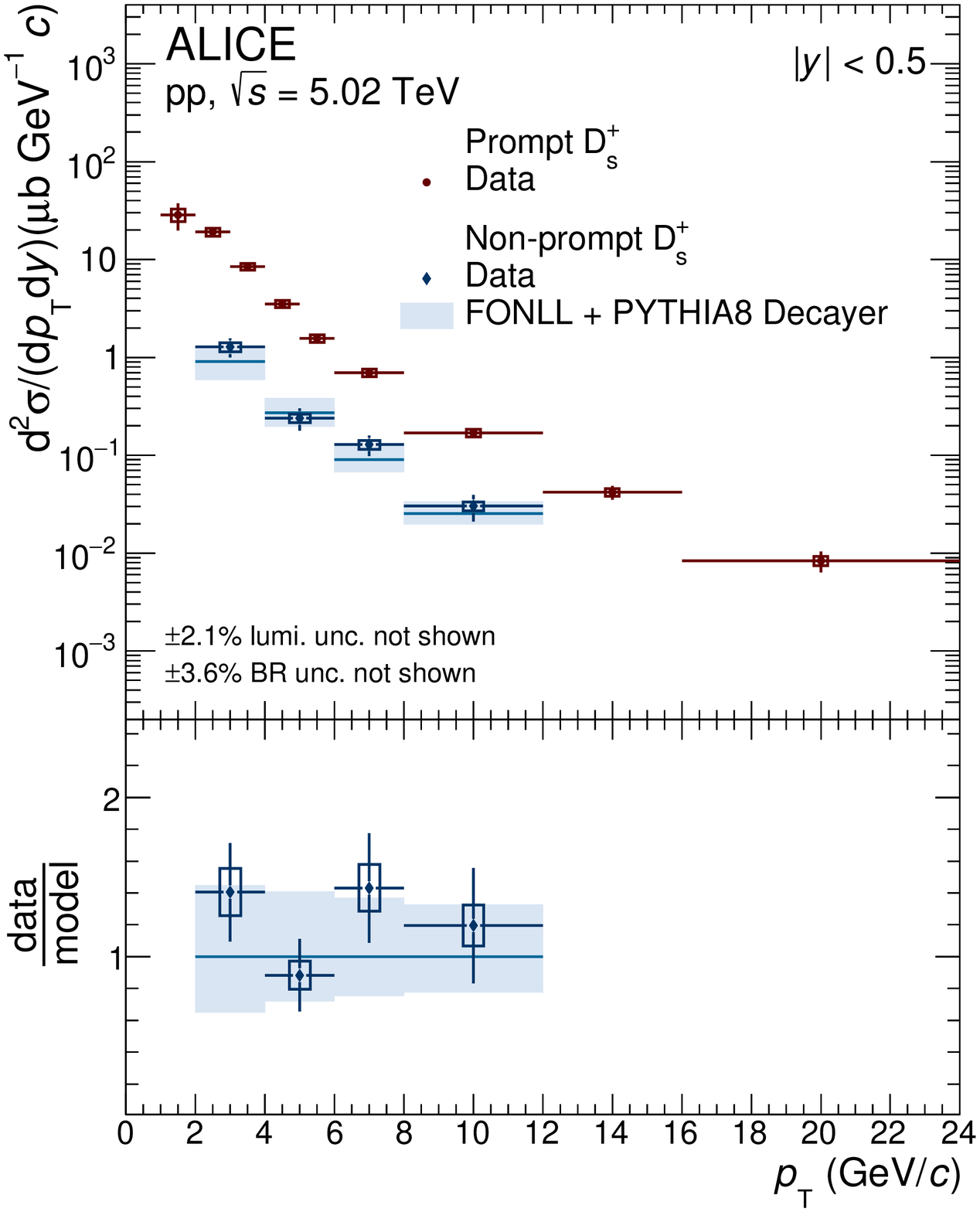}
    \end{center}
    \caption{$\pt$-differential production cross sections of prompt and non-prompt $\Dzero$ (top left panel), $\Dplus$ (top right panel), and $\Ds$ (bottom panel) mesons compared to predictions obtained with FONLL calculations~\cite{Cacciari:1998it,Cacciari:2001td} combined with PYTHIA~8~\cite{Sjostrand:2006za, Sjostrand:2014zea} for the $\bhad\to\mathrm{D+X}$ decay kinematics.  The measurement of prompt $\Dzero$ mesons is the one reported in Ref.~\cite{Acharya:2019mgn}, with updated decay BR as discussed in the text.}
    \label{fig:pt_diff_crosssec_FONLL}
\end{figure}

\begin{figure}[!tb]
    \begin{center}
    \includegraphics[width = 0.48\textwidth]{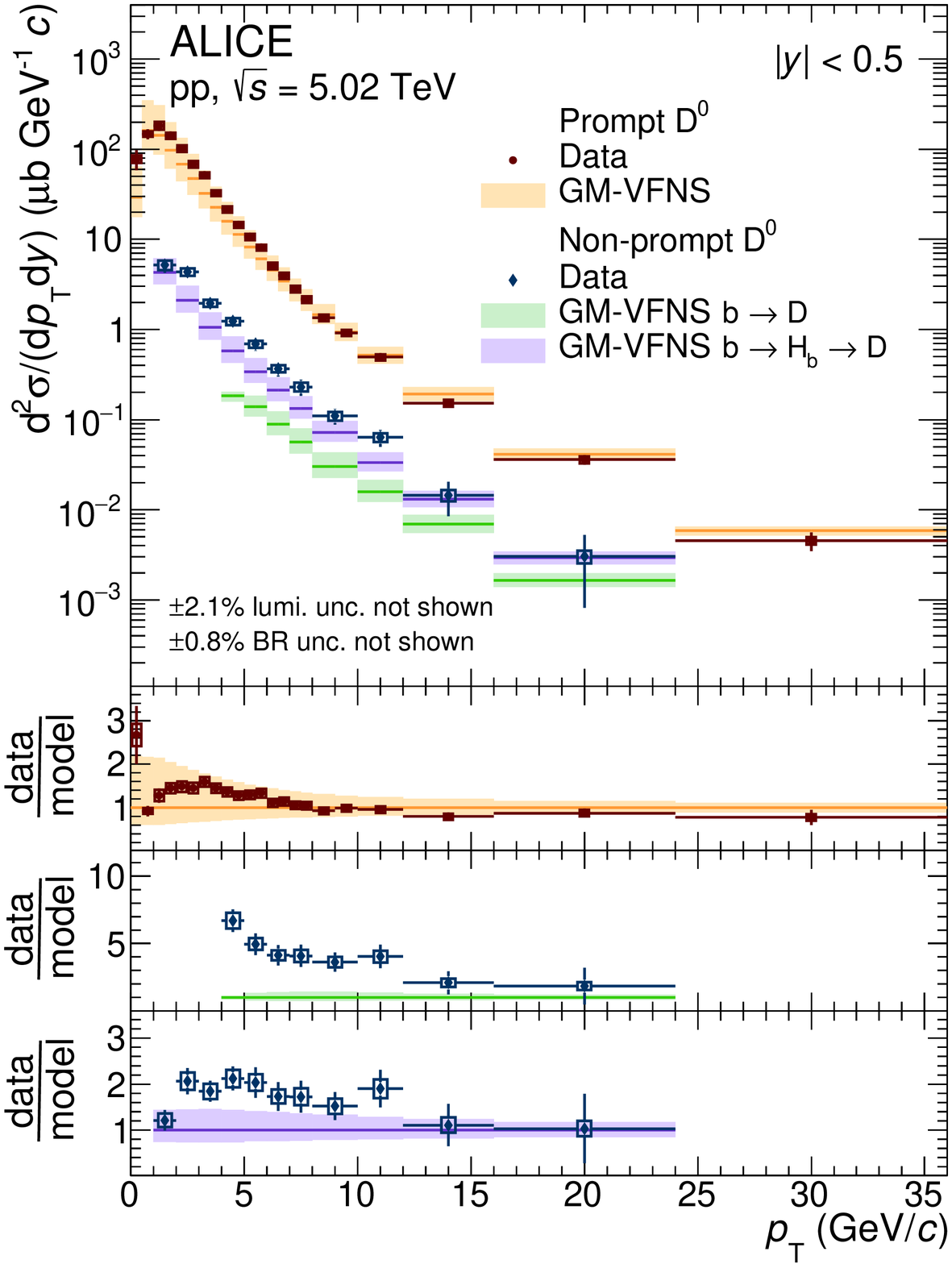}
    \includegraphics[width = 0.48\textwidth]{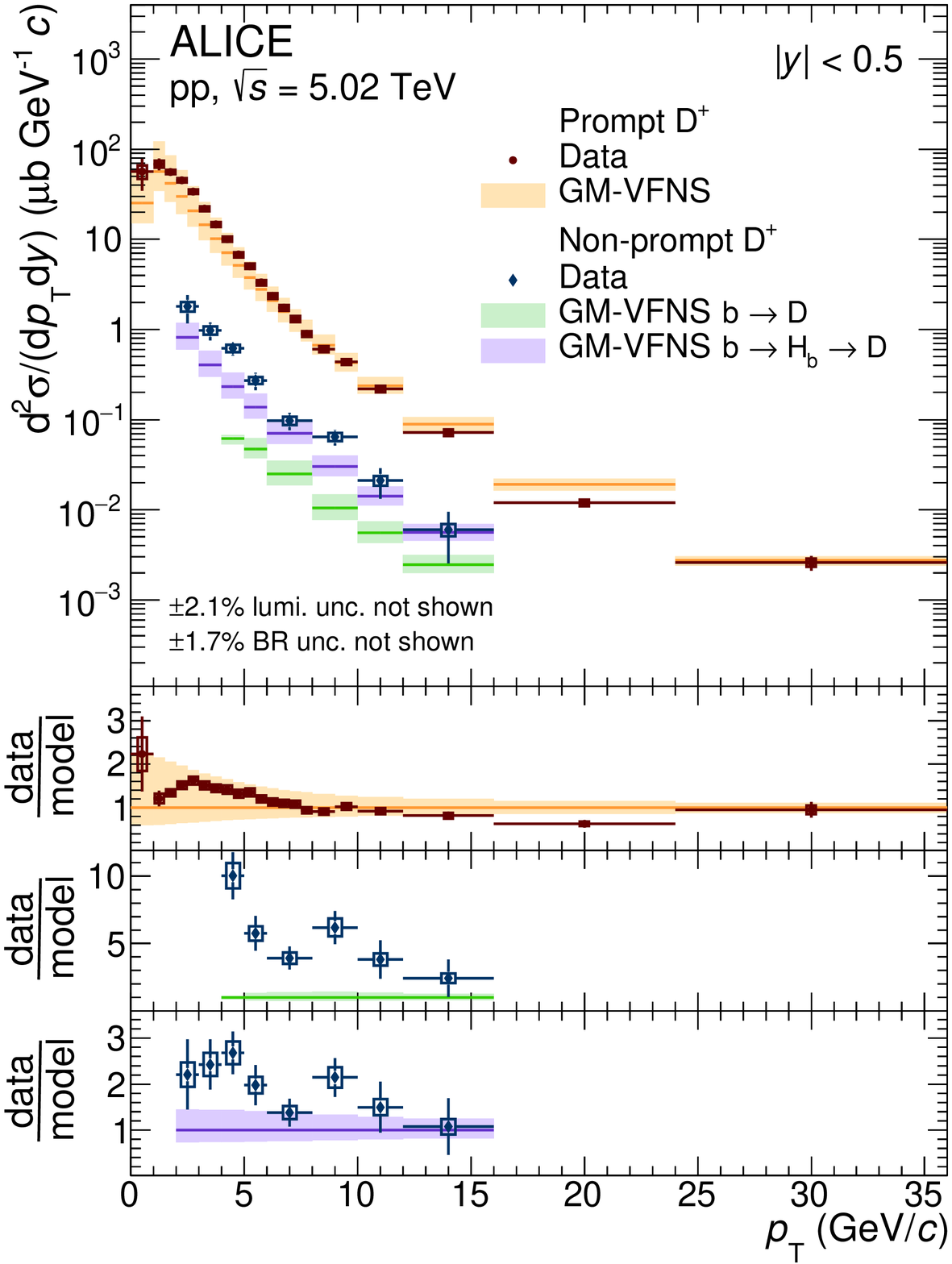}
    \includegraphics[width = 0.48\textwidth]{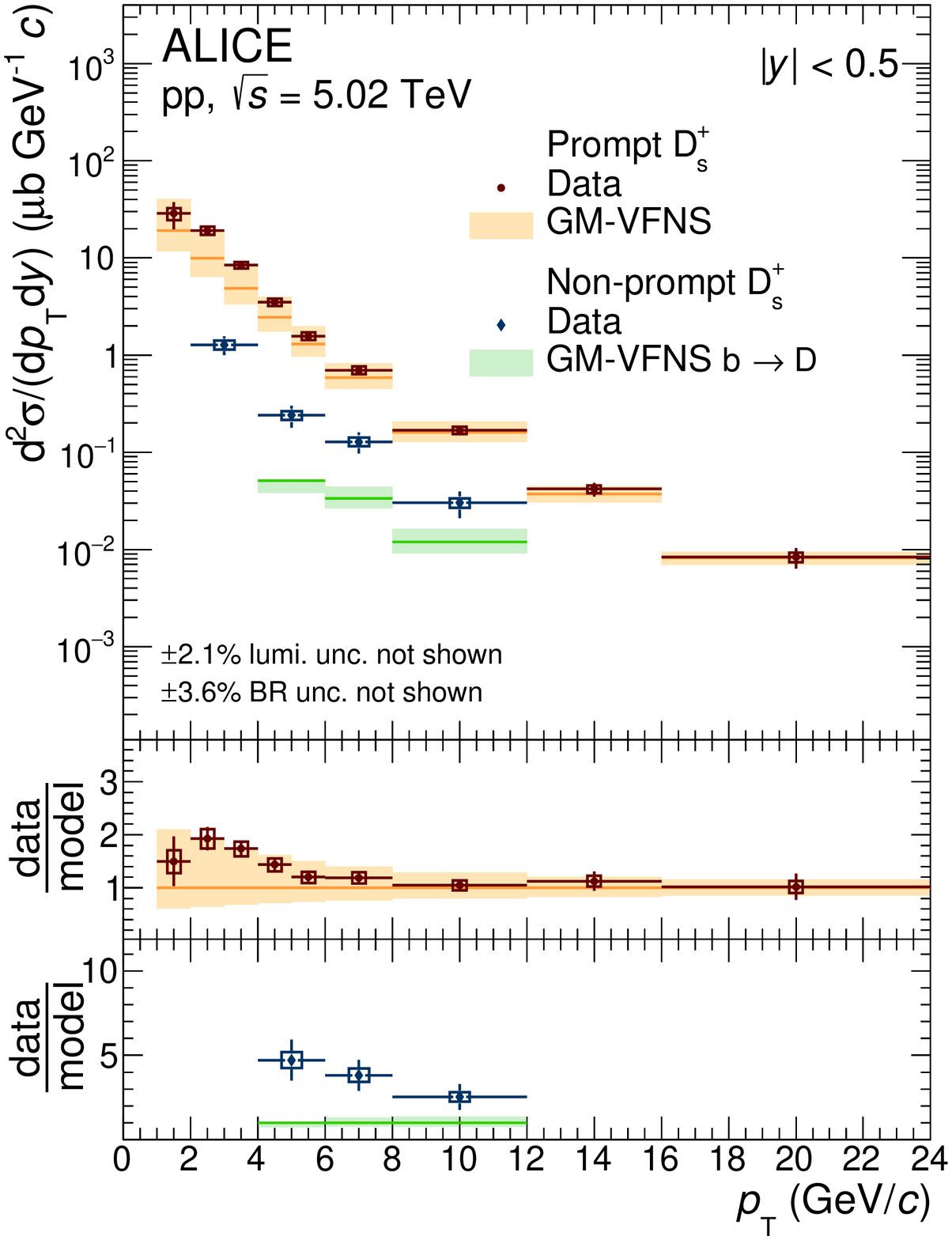}
    \end{center}
    \caption{$\pt$-differential production cross sections of prompt and non-prompt $\Dzero$ (top left panel), $\Dplus$ (top right panel), and $\Ds$ (bottom panel) mesons compared to predictions obtained with GM-VFNS calculations~\cite{Kramer:2017gct,Benzke:2017yjn,Bolzoni:2013vya}. For the non-prompt $\Dzero$ and $\Dplus$ mesons the one-step (green) and two-step (purple) approaches, describing the transition from the beauty quark to the charm meson, are reported. The measurement of prompt $\Dzero$ mesons is the one reported in Ref.~\cite{Acharya:2019mgn}, with updated decay BR as discussed in the text.}
    \label{fig:pt_diff_crosssec_GMVFNS}
\end{figure}

The $\pt$-differential cross sections of prompt and non-prompt D mesons are compared to predictions obtained with FONLL~\cite{Cacciari:1998it,Cacciari:2001td,Cacciari:2012ny} and GM-VFNS~\cite{Kramer:2017gct,Benzke:2017yjn,Bolzoni:2013vya} pQCD calculations in Fig.~\ref{fig:pt_diff_crosssec_FONLL} and Fig.~\ref{fig:pt_diff_crosssec_GMVFNS}, respectively. 
The FONLL uncertainty band includes the uncertainties due to the choice of the renormalisation ($\muR$) and factorisation ($\muF$) scales and of the c and b quark masses, as well as the uncertainties on the CTEQ6.6 PDFs~\cite{Pumplin:2002vw}. In GM-VFNS, the uncertainty related to the choice of the scales is estimated by varying only $\muR$ and the CTEQ14 PDFs~\cite{Dulat:2015mca} are employed. Within the FONLL framework, the fragmentation fractions $\fctoD$ from Ref.~\cite{Gladilin:2014tba} were used  to normalise the prompt $\Dzero$- and $\Dplus$-meson cross sections, while a calculation of the prompt $\Ds$-meson production cross section is not available.
For non-prompt D mesons, FONLL calculations were used to compute the beauty-hadron cross section, while PYTHIA~8~\cite{Sjostrand:2006za, Sjostrand:2014zea} was used for the description of $\bhad\to\mathrm{D+X}$ decay kinematics and branching ratios. The contributions from the different beauty-hadron species were weighted according to fragmentation fractions of b quarks into b-hadron species $\fbtoHb$ measured in the $\Ztobbbar$ decays~\cite{Zyla:2020zbs} reported in Table~\ref{tab:beautyFF}, which provide a good normalisation for B-meson measurements performed by the ATLAS, CMS, and LHCb Collaborations~\cite{ATLAS:2013cia,Aaij:2012jd,Sirunyan:2017oug}.
Two different approaches are instead considered in the GM-VFNS framework. In the first one, the transition from the beauty quark to the charm hadron is described in a single step, exploiting a set of FFs for $\btoDX$ obtained from measurements in $\ee$ collisions as described in Refs.~\cite{Kneesch:2007ey,Kniehl:2006mw}. In the second approach~\cite{Bolzoni:2013vya}, the $\btoDX$ transition is treated in two separate steps, consisting in the $\mathrm{b}\to\bhad$ fragmentation and the $\bhad\to\mathrm{D+X}$ decay, similarly to what was performed in the FONLL+PYTHIA8 calculation. For this latter approach, only predictions for $\Dzero$ and $\Dplus$ mesons are available.

The measured $\pt$-differential cross sections of prompt $\Dzero$, $\Dplus$, and $\Ds$ mesons are described within uncertainties by the FONLL and GM-VFNS predictions. In the case of FONLL, the data lie on the upper edge of the theory uncertainty band, while for the GM-VFNS calculation, the central values of the predictions tend to underestimate the data at low and intermediate $\pt$ and to overestimate them at high $\pt$. The measured non-prompt D-meson cross sections are instead in better agreement with the central values of the FONLL+PYTHIA~8 predictions, while they are underestimated by the GM-VFNS calculations. In the case of the one-step approach, the predictions are lower than the data by a factor ranging between 2 and 10 depending on the $\pt$ and the particle species. The two-step approach describes better the non-prompt $\Dzero$ and $\Dplus$ measurements, nevertheless it still underestimates the measured cross sections.
This confirms that all the different terms of the factorisation approach play a crucial role in the description of the heavy-flavour hadron cross sections, indicating the importance of setting stronger constraints on the fragmentation and decay kinematics.

\begin{table}[!t]
\caption{Fragmentation fractions of b-quarks into beauty-hadron species in $\Ztobbbar$ decays, and in $\ppbar$ collisions at $\s=1.96~\TeV$~\cite{Zyla:2020zbs}.}
\centering
\renewcommand*{\arraystretch}{1.2}
\begin{tabular}[t]{l|>{\centering}p{0.2\linewidth}>{\centering\arraybackslash}p{0.2\linewidth}}
\toprule
b-hadron & Fraction at Z (\%) & Fraction at $\ppbar$ (\%)\\
\midrule
$\Bzero$, $\Bplus$ & $40.8\pm 0.7$ & $34.4\pm 2.1$\\
$\Bs$ & $10.0\pm 0.8$ & $11.5\pm 1.3$ \\
$\Lambdab$ & $8.4\pm 1.1$ & $19.8\pm 4.6$ \\
\bottomrule
\end{tabular}
\label{tab:beautyFF}	
\end{table}

The visible cross sections of prompt and non-prompt D mesons were computed by integrating the measured $\pt$-differential cross sections in the measured $\pt$ range. The results are reported in Table~\ref{tab:visible_crosssec}, where the prompt $\Dzero$-meson cross section is the same as in Ref.~\cite{Acharya:2019mgn}, scaled for the updated BR of the $\DzerotoKpi$ decay channel reported in Ref.~\cite{Zyla:2020zbs}. In the integration of the $\pt$-differential cross sections, the systematic uncertainties were propagated as fully correlated among the measured $\pt$ intervals, except for the raw-yield extraction uncertainty, which was treated as uncorrelated considering the variations of the signal-to-background ratio and the shape of the combinatorial-background distribution as a function of $\pt$. The $\pt$-integrated production cross sections in $|y|<0.5$ were evaluated by multiplying the visible cross sections by an extrapolation factor calculated as follows. For prompt D mesons, the extrapolation factor for each D-meson species was computed using the FONLL central predictions to evaluate the ratio between the production cross section in $|y|<0.5$ and that in the measured $\pt$ interval. The systematic uncertainties on the extrapolation factor were estimated by considering (i) the variation of the factorisation and renormalisation scales in the FONLL calculation, (ii) the uncertainty on the mass of the charm quark, and (iii) the CTEQ6.6 PDFs uncertainties, as proposed in Ref.~\cite{Cacciari:2012ny}. Since FONLL predictions are not available for prompt $\Ds$ mesons, the central value of the extrapolation factor was computed as described in Ref.~\cite{Acharya:2019mgn}, using the prediction based on the $\pt$-differential cross section of charm quarks from FONLL, the fragmentation fractions $f(\mathrm{c}\rightarrow\Ds)$ and $f(\mathrm{c}\rightarrow\Dsstar)$ from ALEPH measurements~\cite{Barate:1999bg}, and the charm fragmentation functions from Ref.~\cite{Braaten:1994bz}.
The measurements of $\Dzero$ and $\Dplus$ mesons extend from $\pt = 0$ up to $\pt=36~\GeV/c$, leading to an extrapolation factor close to unity and a negligible associated uncertainty. In the case of non-prompt D mesons, the extrapolation factor was evaluated using the FONLL predictions for the beauty-hadron production and PYTHIA~8 to describe the $\bhad\to\mathrm{D+X}$ decay kinematics. Besides the uncertainties of FONLL, for the non-prompt D-meson extrapolation factors two additional sources of systematic uncertainties were considered, i.e. the uncertainty on (i) the beauty fragmentation fractions $\fbtoHb$ and (ii) the branching ratios of the $\bhad\to\mathrm{D+X}$ decays. The former was estimated considering an alternative set of beauty fragmentation fractions measured in $\ppbar$ collisions~\cite{Zyla:2020zbs} reported in Table~\ref{tab:beautyFF}, while for the latter the branching ratios implemented in PYTHIA~8 were reweighted in order to reproduce the measured values reported in Ref.~\cite{Zyla:2020zbs}. In addition, it was verified that the extrapolation factors computed with the PYTHIA~8 decayer were compatible with those resulting from the usage of the EvtGen package~\cite{Lange:2001uf} for the description of the beauty-hadron decays. The production cross sections for prompt and non-prompt D mesons in $|y|<0.5$ are reported in Table~\ref{tab:ptint_crosssec}.
The cross sections of prompt $\Ds$ and $\Dplus$ mesons are compatible with those reported in Ref.~\cite{Acharya:2019mgn}, but their total uncertainties are reduced, owing to the improved precision of the $\pt$-differential measurements and the extended $\pt$ range, which implies a smaller fraction of extrapolated cross section.

\begin{table}[!tb]
\caption{$\pt$-integrated production cross sections in the measured $\pt$ range for prompt and non-prompt D mesons in the range $|y|<0.5$ in pp collisions at $\s=5.02~\TeV$.}
\centering
\renewcommand*{\arraystretch}{1.2}
\begin{tabular}[t]{l|>{\centering}p{0.3\linewidth}>{\centering\arraybackslash}p{0.48\linewidth}}
\toprule
Meson & Kinematic range $(\GeV/c)$ & Visible cross section ($\mub$)\\
\midrule
Prompt & \multicolumn{2}{c}{} \\
\midrule
$\Dzero$ & $0<\pt<36$ & $440 \pm 19 ({\rm stat}) \pm 29 ({\rm syst}) \pm 9 ({\rm lumi}) \pm 3 ({\rm BR})$\\
$\Dplus$ & $0<\pt<36$ &  $194 \pm 23 ({\rm stat}) \pm 16 ({\rm syst}) \pm 4 ({\rm lumi}) \pm 3 ({\rm BR})$\\
$\Ds$ & $1<\pt<24$ & $64 \pm 9 ({\rm stat}) ^{+6}_{-7} ({\rm syst}) \pm 1 ({\rm lumi}) \pm 2 ({\rm BR})$\\
\midrule
Non-prompt & \multicolumn{2}{c}{} \\
\midrule
$\Dzero$ & $1<\pt<24$ & $14.5 \pm 1.2 (\mathrm{stat}) \pm 1.3 (\mathrm{syst}) \pm 0.3 (\mathrm{lumi}) \pm 0.1 (\mathrm{BR})$\\
$\Dplus$ &  $2<\pt<16$ & $4.1 \pm 0.7 (\mathrm{stat}) \pm 0.4 (\mathrm{syst}) \pm 0.1 (\mathrm{lumi}) \pm 0.1 (\mathrm{BR})$\\
$\Ds$ & $2<\pt<12$ & $3.4 \pm 0.6 (\mathrm{stat}) \pm 0.3 (\mathrm{syst}) \pm 0.1 (\mathrm{lumi}) \pm 0.1 (\mathrm{BR})$\\
\bottomrule
\end{tabular}
\label{tab:visible_crosssec}	
\end{table}

\begin{table}[!tb]
\caption{Production cross sections of prompt and non-prompt D mesons in the range $|y| < 0.5$ in pp collisions at $\s=5.02~\TeV$.}
\centering
\renewcommand*{\arraystretch}{1.2}
\begin{tabular}[t]{l|>{\centering}p{0.21\linewidth}>{\centering\arraybackslash}p{0.59\linewidth}}
\toprule
Meson & Extr. factor to $\pt>0$ & $\de\sigma/\de y |_{|y|<0.5}$ ($\mub$)\\
\midrule
Prompt & \multicolumn{2}{c}{} \\
\midrule
$\Dzero$ & $1.0000^{+0.0003}_{-0.0000}$ & $440 \pm 19 ({\rm stat}) \pm 29 ({\rm syst}) \pm 9 ({\rm lumi}) \pm 3 ({\rm BR})$\\
$\Dplus$ & $1.0000^{+0.0003}_{-0.0000}$ & $195 \pm 23 ({\rm stat}) \pm 16 ({\rm syst}) \pm 4 ({\rm lumi}) \pm 3 ({\rm BR})$\\
$\Ds$ & $1.28^{+0.35}_{-0.12}$ & $82 \pm 12 ({\rm stat}) \pm 8 ({\rm syst}) \pm 2 ({\rm lumi}) \pm 3 ({\rm BR})^{+23}_{-8} ({\rm extr})$\\
\midrule
Non-prompt & \multicolumn{2}{c}{} \\
\midrule
$\Dzero$ & $1.28^{+0.01}_{-0.04}$ & $18.4 \pm 1.5 (\mathrm{stat}) \pm 1.6 (\mathrm{syst}) \pm 0.4 (\mathrm{lumi}) \pm 0.1 (\mathrm{BR}) ^{+0.1}_{-0.6} (\mathrm{extr})$\\
$\Dplus$ &  $2.22^{+0.05}_{-0.19}$ & $9.0 \pm 1.5 (\mathrm{stat}) \pm 0.9 (\mathrm{syst}) \pm 0.2 (\mathrm{lumi}) \pm 0.2 (\mathrm{BR}) ^{+0.2}_{-0.8} (\mathrm{extr})$\\
$\Ds$ & $2.03^{+0.04}_{-0.15}$ & $6.9 \pm 1.2 (\mathrm{stat}) \pm 0.7 (\mathrm{syst}) \pm 0.1 (\mathrm{lumi}) \pm 0.2 (\mathrm{BR}) ^{+0.1}_{-0.5} (\mathrm{extr})$\\
\bottomrule
\end{tabular}
\label{tab:ptint_crosssec}	
\end{table}

\subsection{Cross section ratios}
\label{sec:resultsRatios}

The $\pt$-integrated cross sections were used to compute the ratios of production yields among the different D-meson species reported in Table~\ref{tab:ptint_crosssecratios}. In the computation of these ratios, the systematic uncertainties related to the tracking efficiency, luminosity, and, for the prompt D mesons, the contribution due to the subtraction of the component from beauty-hadron decays, were considered as correlated among the different D-meson species. The extrapolation uncertainties were also treated as correlated, except for the source of uncertainty due to the branching ratios of the beauty-hadron decays used in the extrapolation of the $\pt$-integrated cross section of non-prompt D mesons. All the other sources of systematic uncertainties were propagated as uncorrelated. The $\Dplus/\Dzero$ ratio is compatible between prompt and non-prompt D-meson production, while for the $\Ds$ over non-strange D meson ratios, the measured values are higher for non-prompt D mesons than for prompt D mesons with a significance of about $2.5\,\sigma$. This finding is qualitatively expected from the $\mathrm{b\rightarrow{c\overline{c}s}}$ and $\mathrm{\overline{b}\rightarrow{c\overline{c}\overline{s}}}$ weak decays, which enhance $\Ds$ final states. Moreover, it is consistent with previous measurements at LEP~\cite{Gladilin:2014tba}.

\begin{table}[!tb]
\caption{Ratios of the measured production cross sections of prompt and non-prompt D mesons in the range $|y| < 0.5$ in pp collisions at $\s=5.02~\TeV$.}
\centering
\renewcommand*{\arraystretch}{1.4}
\begin{tabular}[t]{l|>{\centering\arraybackslash}p{0.65\linewidth}}
\toprule
\multicolumn{2}{c}{Prompt}\\
\midrule
$\Dplus/\Dzero$ & $ 0.442 \pm 0.055(\mathrm{stat})\pm0.033(\mathrm{syst}) \pm0.008(\mathrm{BR})$\\
$\Ds/\Dzero$ & $0.186 \pm 0.028(\mathrm{stat})\pm 0.015(\mathrm{syst}) \pm 0.007(\mathrm{BR}) ^{+0.051}_{-0.018}(\mathrm{extr})$\\
$\Ds/\Dplus$ & $0.419 \pm 0.078(\mathrm{stat})\pm0.041(\mathrm{syst}) \pm0.017(\mathrm{BR}) ^{+0.116}_{-0.040}(\mathrm{extr})$\\
$\Ds/(\Dzero+\Dplus)$ & $ 0.128 \pm 0.020(\mathrm{stat})\pm0.010(\mathrm{syst}) ^\pm0.005(\mathrm{BR}) ^{+0.035}_{-0.012}(\mathrm{extr})$\\
\midrule
\multicolumn{2}{c}{Non-prompt} \\
\midrule
$\Dplus/\Dzero$ & $0.487 \pm 0.090(\mathrm{stat}) \pm0.055(\mathrm{syst}) \pm0.009(\mathrm{BR}) ^{+0.007}_{-0.027}(\mathrm{extr})$\\
$\Ds/\Dzero$ & $0.375 \pm 0.071(\mathrm{stat})\pm0.041(\mathrm{syst})\pm0.014(\mathrm{BR}) ^{+0.004}_{-0.016}(\mathrm{extr})$\\
$\Ds/\Dplus$ & $0.769 \pm 0.183(\mathrm{stat})\pm0.086(\mathrm{syst})\pm0.030(\mathrm{BR}) ^{+0.003}_{-0.010}(\mathrm{extr})$\\
$\Ds/(\Dzero+\Dplus)$ & $0.252 \pm 0.047(\mathrm{stat})\pm0.023(\mathrm{syst})\pm0.009(\mathrm{BR}) ^{+0.001}_{-0.006}(\mathrm{extr})$\\
\bottomrule
\end{tabular}
\label{tab:ptint_crosssecratios}	
\end{table}

\begin{figure}[!tb]
    \begin{center}
    \includegraphics[width = 0.48\textwidth]{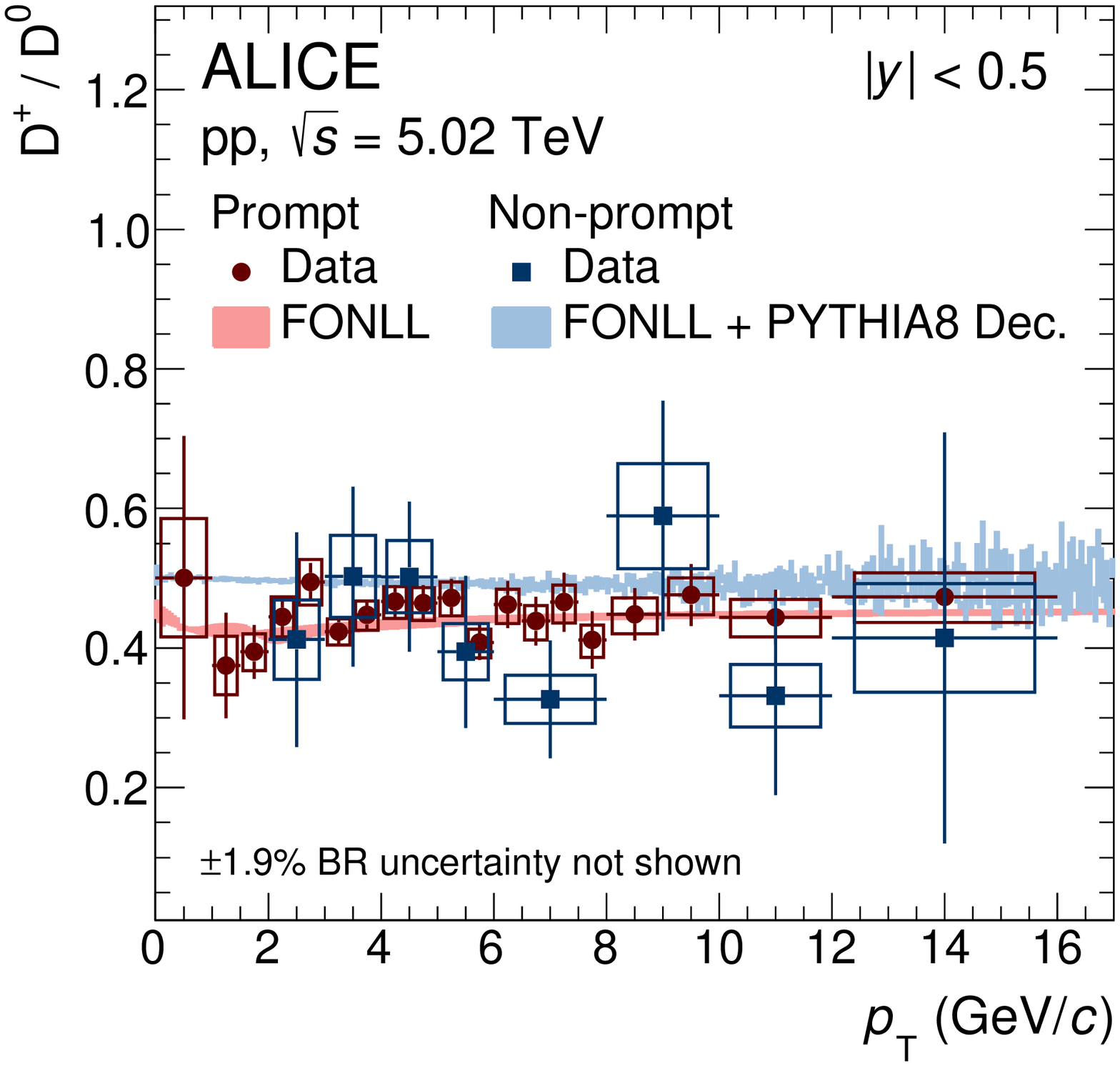}
    \includegraphics[width = 0.48\textwidth]{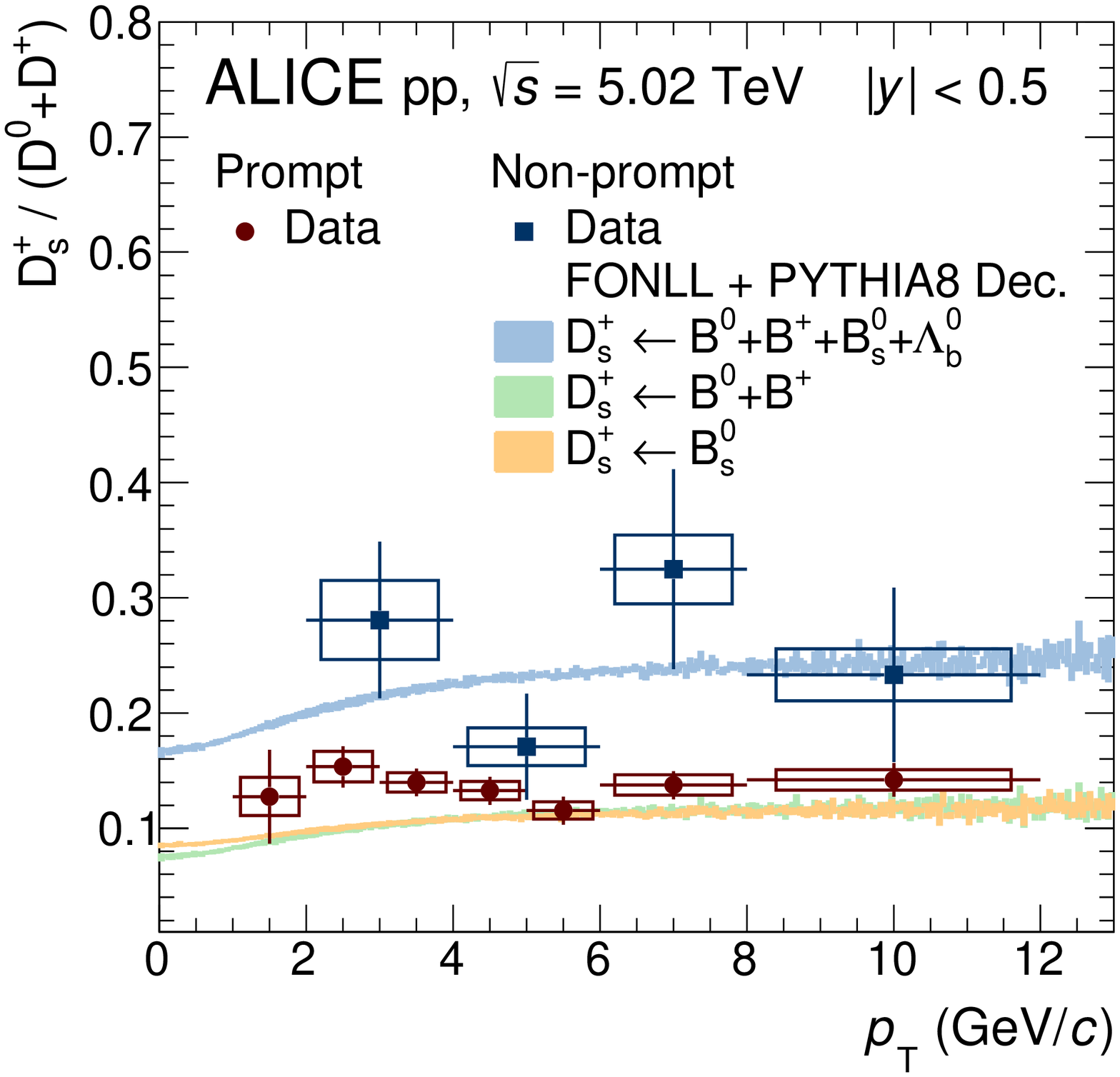}
    \end{center}
    \caption{Ratios between the $\pt$-differential production cross sections of $\Dplus$ and $\Dzero$ mesons (left panel) and between the one of $\Ds$ mesons and the sum of the $\Dzero$- and $\Dplus$-meson cross sections (right panel) compared with predictions obtained with FONLL calculations~\cite{Cacciari:1998it,Cacciari:2001td} and PYTHIA~8~\cite{Sjostrand:2006za, Sjostrand:2014zea} for the $\bhad\to\mathrm{D+X}$ decay kinematics. For the non-prompt $\Ds/(\Dzero+\Dplus)$ ratio, the predictions for the $\Ds$ from $\Bs$ and from non-strange B meson decays are also displayed separately.}
    \label{fig:d_meson_ratios}
\end{figure}

A possible $\pt$ dependence was investigated computing the $\pt$-differential ratios. The ratios between the $\pt$-differential production cross sections of $\Dplus$ and $\Dzero$ mesons and the ratios between the one of $\Ds$ mesons and the sum of the $\Dzero$ and $\Dplus$ mesons are reported in the left and right panels of Fig.~\ref{fig:d_meson_ratios}, respectively. The measured ratios are independent of $\pt$ in the measured $\pt$ range within the current experimental precision. They are also compatible with the FONLL predictions in the case of prompt $\Dzero$ and $\Dplus$ mesons and FONLL+PYTHIA~8 in the case of non-prompt D mesons. In the right panel of Fig.~\ref{fig:d_meson_ratios}, the contributions of $\Ds$ from $\Bs$ and non-strange B meson decays in the FONLL+PYTHIA~8 calculation are depicted separately to highlight the substantial contribution of non-prompt $\Ds$ mesons from the decay of non-strange B mesons.

The prompt $\Ds/(\Dzero+\Dplus)$ ratio represents the fragmentation fraction of charm quarks to charm-strange mesons $\f{s}$ divided by the one to non-strange charm mesons $\f{u}+\f{d}$, given that all $\Dstar$ and $\DstarZero$ mesons decay to $\Dzero$ and $\Dplus$ mesons, and all $\DstarS$ mesons decay to $\Ds$ mesons. Considering that the uncertainties in the production ratios reported in Table~\ref{tab:ptint_crosssecratios} are dominated by the limited precision of the measurements in the low $\pt$ region and that the $\pt$-differential ratios are constant within uncertainties, the ratio of charm-quark fragmentation fractions was computed by fitting the data with a constant function, leading to
\begin{equation}
\bigg{(}\frac{\f{s}}{\f{u}+\f{d}}\bigg{)}_\mathrm{charm} = 0.136 \pm 0.005(\mathrm{stat})\pm 0.006(\mathrm{syst}) \pm 0.005(\mathrm{BR}).
\end{equation}
In addition to the degree of correlation among the D-meson species considered for the computation of the $\pt$-differential ratios, all the sources of systematic uncertainties except for the one related to the raw-yield extraction were propagated as fully correlated among the different $\pt$ intervals. A similar strategy was adopted by the LHCb Collaboration for the beauty sector in Ref.~\cite{Aaij:2011jp}.

\begin{figure}[!tb]
    \begin{center}
    \includegraphics[width = 0.75\textwidth]{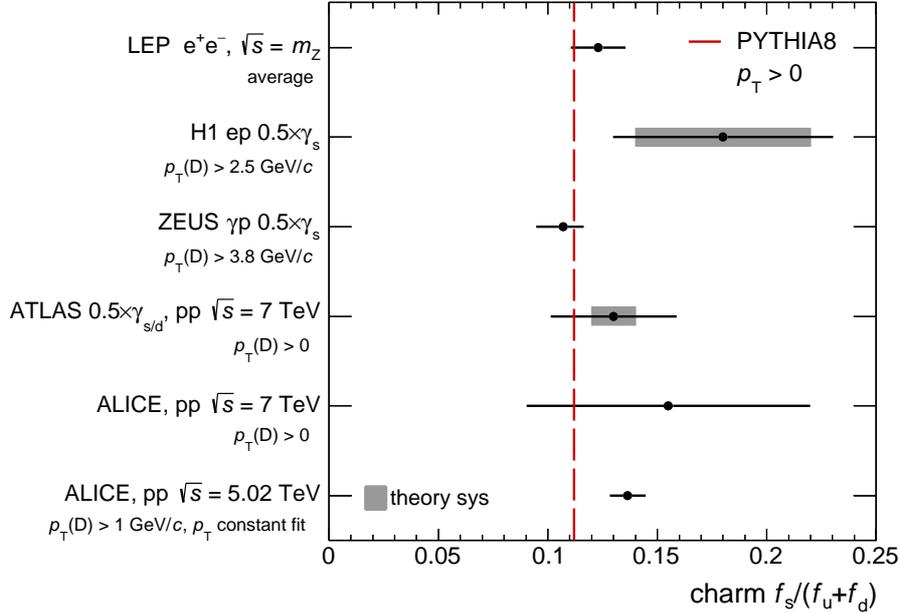}
    \end{center}
    \caption{Charm-quark fragmentation-fraction ratio $\f{s}/(\f{u}+\f{d})$ compared with previous measurements performed by the ALICE~\cite{Abelev:2012tca}, H1~\cite{Aktas:2004ka}, ZEUS~\cite{Abramowicz:2013eja}, and ATLAS~\cite{Aad:2015zix} Collaborations and to the average of LEP measurements~\cite{Gladilin:2014tba}. The total experimental uncertainties (bars) and the theoretical uncertainties (shaded boxes) are shown. The experimental measurements are compared to the value obtained from PYTHIA~8 simulations with Monash-13 tune ~\cite{Skands:2014pea}.}
    \label{fig:charm_fs}
\end{figure}

In Fig.~\ref{fig:charm_fs}, the charm-quark fragmentation-fraction ratio $\f{s}/(\f{u}+\f{d})$ is compared with previous measurements of strangeness suppression factor $\gamma_\mathrm{s}$ from the ALICE~\cite{Abelev:2012tca}, H1~\cite{Aktas:2004ka}, ZEUS~\cite{Abramowicz:2013eja}, and ATLAS~\cite{Aad:2015zix} Collaborations. They were divided by a factor two to account for the difference between $\gamma_\mathrm{s}$ and the ratio of fragmentation fractions $\f{s}/(\f{u}+\f{d})$. The theoretical uncertainties in case of the H1 result include the branching ratio uncertainty and the model dependencies of the acceptance determination, while for the ATLAS result the extrapolation uncertainties to the full phase space are included. All the values are compatible within uncertainties and with the average of measurements at LEP~\cite{Gladilin:2014tba}. The experimental points are also compared to the value obtained from PYTHIA~8 simulations with Monash-13 tune~\cite{Skands:2014pea} and found to be compatible with it within the uncertainties, even if a tension of about 2.7 standard deviations (including both statistical and systematic uncertainties) is observed for the result presented in this paper.

A similar procedure was followed to obtain the fragmentation fraction of beauty quarks to beauty-strange mesons divided by the one to non-strange beauty mesons, starting from the measured non-prompt $\Ds/(\Dzero+\Dplus)$ ratio. In the case of non-prompt D mesons, an additional correction factor was necessary to account for the fraction of non-prompt $\Ds$ mesons not originating from $\Bs$ decays and that of non-prompt $\Dzero$ and $\Dplus$ mesons not originating from non-strange B-meson decays. This correction factor was computed from FONLL+PYTHIA~8 and a systematic uncertainty was assigned by varying the set of beauty fragmentation fractions and the beauty-hadron branching ratios, as described in Section~\ref{sec:NPD_cross_sec}. In the case of $\Ds$ mesons, $\Bs$ and non-strange B mesons are expected to contribute almost equally to the non-prompt $\Ds$ cross section as shown in the right panel of Fig.~\ref{fig:d_meson_ratios}, while most of the non-prompt $\Dzero$ and $\Dplus$ mesons come from non-strange B-meson decays. The $\pt$-differential ratio of beauty-quark fragmentation fractions was then computed as
\begin{equation}
\bigg{(}\frac{\f{s}}{\f{u}+\f{d}}\bigg{)}_\mathrm{beauty} = \bigg{[}\frac{N(\Ds\leftarrow\Bs)}{N(\Ds\leftarrow\bhad)}\times\frac{N(\Dzero,\Dplus\leftarrow\bhad)}{N(\Dzero,\Dplus\leftarrow\Bzeroplus)}\bigg{]}^\mathrm{FONLL+PYTHIA~8}\times\bigg{(}\frac{\Ds}{\Dzero+\Dplus}\bigg{)}_\mathrm{non-prompt},
\label{eq:fsfufd_beauty}
\end{equation}
and fitted with a constant function, as done for the prompt D mesons. The result is 
\begin{equation}
\bigg{(}\frac{\f{s}}{\f{u}+\f{d}}\bigg{)}_\mathrm{beauty} =  0.127 \pm 0.036(\mathrm{stat})\pm 0.012(\mathrm{syst}) \pm 0.005(\mathrm{BR}) \pm 0.005(\mathrm{th}),
\end{equation}
where the theoretical uncertainty arises from the correction factor in Eq.~\ref{eq:fsfufd_beauty} for the fractions of $\Ds$ ($\Dzero$ and $\Dplus$) mesons originating from $\Bs$($\Bzeroplus$)-meson decays. 

\begin{figure}[!tb]
    \begin{center}
    \includegraphics[width = 0.75\textwidth]{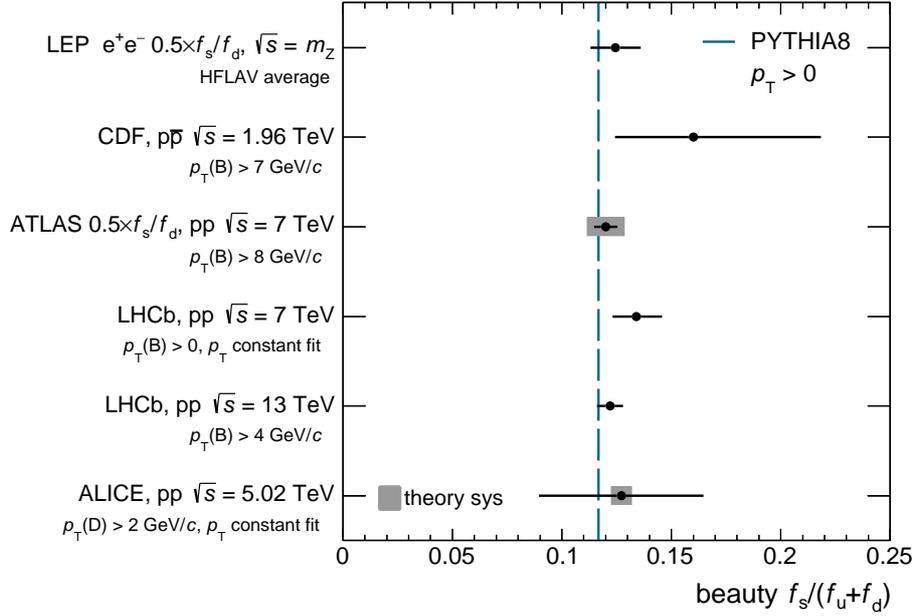}
    \end{center}
    \caption{Beauty-quark fragmentation-fraction ratio $\f{s}/(\f{u}+\f{d})$ from non-prompt D-meson measurements compared with previous measurements performed by the CDF~\cite{Aaltonen:2008zd}, LHCb~\cite{Aaij:2011jp,Aaij:2019pqz}, and ATLAS~\cite{Aad:2015cda} Collaborations and to the average of LEP measurements~\cite{Amhis:2019ckw}. The total experimental uncertainties (bars) and the theoretical uncertainties (shaded boxes) are shown. The experimental measurements are compared to the value obtained from PYTHIA~8 simulations with Monash-13 tune ~\cite{Skands:2014pea}.}
    \label{fig:beauty_fs}
\end{figure}

The beauty-quark fragmentation-fraction ratio $\f{s}/(\f{u}+\f{d})$ is compared with previous measurements from CDF~\cite{Aaltonen:2008zd}, LHCb~\cite{Aaij:2011jp,Aaij:2019pqz}, and ATLAS~\cite{Aad:2015cda} Collaborations in Fig.~\ref{fig:beauty_fs}. The ATLAS measurement was divided by a factor two assuming isospin symmetry for the u and d quarks, which implies $\f{u}=\f{d}$. All the $\f{s}/(\f{u}+\f{d})$ values measured in pp and $\ppbar$ collisions are found to be compatible with the LEP average, computed by the HFLAV Collaboration~\cite{Amhis:2019ckw} and the value obtained from PYTHIA~8 simulations with Monash-13 tune~\cite{Skands:2014pea}. It is also interesting to note that the fragmentation-fraction ratios $\f{s}/(\f{u}+\f{d})$ are similar for the charm and beauty sectors and are consistent with the ratio of light strange to non-strange particle production in pp and $\ee$ collisions~\cite{BraunMunzinger:2001as}.

\subsection{Extrapolation to the $\pmb{\bbbar}$ production cross section}
\label{sec:resultsBeauty}
The $\bbbar$ production cross section per unit of rapidity at midrapidity ($|y|<0.5$) was computed following a similar procedure as the one adopted to derive the $\pt$-integrated production cross sections of non-prompt D mesons. In this case, the extrapolation factor $\alpha_\mathrm{extr}^\bbbar$ was computed as
\begin{equation}
    \alpha_\mathrm{extr}^\bbbar = \frac{\de\sigma_\bbbar/\de y|_{|y|<0.5}^\mathrm{FONLL}}{\sigma_\mathrm{b\to D}^\mathrm{FONLL+PYTHIA~8}(\pt^\mathrm{min}<\pt<\pt^\mathrm{max}, |y|<0.5)},
\label{eq:extrap_bbbar}
\end{equation}
where $\de\sigma_\bbbar/\de y|_{|y|<0.5}^\mathrm{FONLL}$ is the $\bbbar$ production cross section obtained with FONLL calculations with a correction for the different shapes of the rapidity distributions of beauty hadrons and $\bbbar$ pairs, and $\sigma_\mathrm{b\to D}^\mathrm{FONLL+PYTHIA~8}(\pt^\mathrm{min}<\pt<\pt^\mathrm{max}, |y|<0.5)$ is the non-prompt D meson cross section in the measured phase space from the FONLL+PYTHIA~8 model. 
The correction for the $\bbbar$ rapidity distribution is composed of two factors. The first factor accounts for the different rapidity distributions of beauty mesons and single beauty quarks and it was evaluated to be unity in the relevant rapidity range based on FONLL calculations. A 1\% uncertainty on this factor was evaluated from the difference between values from FONLL and PYTHIA~8. The second correction factor is the ratio $(\de\sigma_\bbbar/\de y) / (\de\sigma_\mathrm{b}/\de y)$, which was estimated from NLO pQCD calculations (POWHEG~\cite{Frixione:2007nw}) as
$\de\sigma_\bbbar^{|y|<0.5}/\de\sigma_\mathrm{b}^{|y|<0.5} = 1.06$. A 1\% uncertainty on this factor was estimated from the difference among the values obtained varying the factorisation and renormalisation scales in the POWHEG calculation and using different sets of PDFs (CT10NLO~\cite{Lai:2010vv} and CT14NLO~\cite{Dulat:2015mca}).
The other sources of systematic uncertainty on the extrapolation factor, i.e. FONLL, BR($\bhad\to\mathrm{D+X}$), and $\fbtoHb$, are the same as those described in Section~\ref{sec:NPD_cross_sec} for the extrapolation of the $\pt$-integrated production cross sections of non-prompt D meson.

The $\de\sigma_\bbbar/\de y$ was computed separately for each D-meson species and the three values were then averaged using the inverse of the quadratic sum of the absolute statistical and uncorrelated systematic uncertainties as weights. The systematic uncertainties related to the tracking uncertainty and the extrapolation uncertainties related to FONLL and the beauty fragmentation fractions were treated as fully correlated among the three D-meson species, while all the other sources as uncorrelated. The resulting $\bbbar$ cross section at midrapidity is
\begin{equation}
    \frac{\de\sigma_\bbbar}{\de y}\bigg{|}_{|y|<0.5} = 34.5 \pm 2.4 (\mathrm{stat}) \pm 2.5 (\mathrm{syst}) \pm 0.7 (\mathrm{lumi}) \pm 0.3 (\mathrm{BR}) ^{+3.8}_{-1.1} (\mathrm{extr}) \pm 0.5 (\mathrm{rap.~shape})~\mub.
\end{equation}

\begin{figure}[!tb]
    \begin{center}
    \includegraphics[width = 0.75\textwidth]{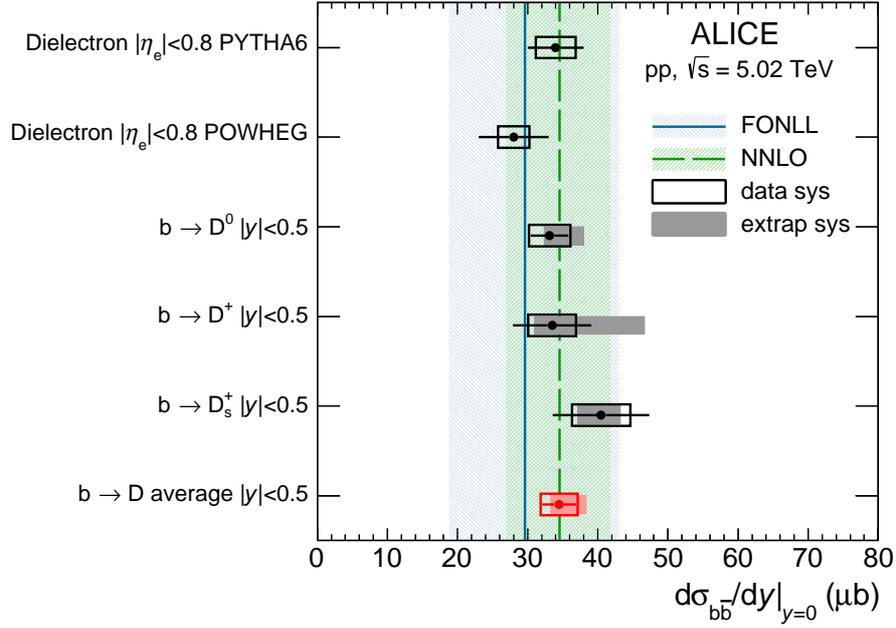}
    \end{center}
    \caption{Estimates of $\de\sigma_\bbbar/\de y$ at midrapidity from dielectron~\cite{Acharya:2020rfm} and non-prompt $\Dzero$, $\Dplus$, and $\Ds$ meson measured in pp collisions at $\s=5.02~\TeV$ compared to FONLL~\cite{Cacciari:1998it,Cacciari:2001td,Cacciari:2012ny} and NNLO~\cite{Catani:2020kkl} predictions. The average $\de\sigma_\bbbar/\de y$ of the estimates from the single D-meson species is also reported.}
    \label{fig:bbbar_crosssec_compilation}
\end{figure}

\begin{figure}[!tb]
    \begin{center}
    \includegraphics[width = 0.75\textwidth]{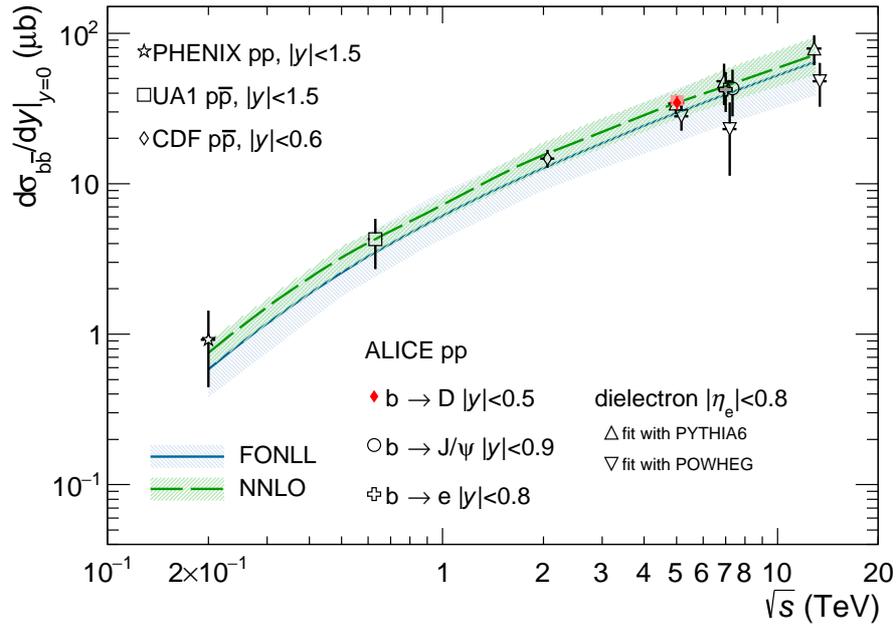}
    \end{center}
    \caption{Beauty production cross section per rapidity unit at midrapidity as a function of $\s$ as measured in pp collisions by the ALICE~\cite{Abelev:2012gx,Abelev:2012sca,Acharya:2018ohw,Acharya:2018kkj} and PHENIX~\cite{Adare:2009ic} Collaborations and in $\ppbar$ collisions by the CDF~\cite{Acosta:2004yw} and UA1~\cite{Albajar:1990zu} Collaborations. The ALICE data points are shifted in the $\s$-axis for better visibility. The solid and dashed lines with the shaded band represents the FONLL~\cite{Cacciari:1998it,Cacciari:2001td,Cacciari:2012ny} and NNLO~\cite{Catani:2020kkl} calculations with their uncertainties, respectively.}
    \label{fig:bbbar_crosssec_vs_sqrt}
\end{figure}

Figure~\ref{fig:bbbar_crosssec_compilation} shows the extrapolated $\de\sigma_\bbbar/\de y$ from each D-meson species and their average, compared to those obtained from dielectron~\cite{Acharya:2020rfm} along with a comparison to FONLL and NNLO calculations. The values extracted from the three D-meson species are compatible within uncertainties among each other and with those obtained from the other two ALICE measurements, as well as with the FONLL and NNLO predictions. As compared to FONLL calculations, the inclusion of NNLO corrections leads to a slightly larger central value, more in agreement with the experimental result based on non-prompt D mesons, and to reduced theoretical uncertainties. The measurements in pp collisions at $\s=5.02~\TeV$ are also shown in Fig.~\ref{fig:bbbar_crosssec_vs_sqrt} along with the other existing measurements in pp collisions by the ALICE ~\cite{Abelev:2012gx,Abelev:2012sca,Acharya:2018ohw,Acharya:2018kkj} and PHENIX~\cite{Adare:2009ic} Collaborations at different centre-of-mass energies, and in $\ppbar$ collisions by the CDF~\cite{Acosta:2004yw} and UA1~\cite{Albajar:1990zu} Collaborations. The experimental results are found to be compatible with FONLL and NNLO calculations. 

\section{Summary}
The $\pt$-differential cross sections of prompt and non-prompt $\Dzero$,  $\Dplus$, and $\Ds$ mesons were measured at midrapidity ($|y|<0.5$) in pp collisions at $\s=5.02~\TeV$ using a machine-learning technique based on Boosted Decision Trees. A data-driven method was employed for the evaluation of the fraction of non-prompt D mesons, $\fnonprompt$, and for the validation of the FONLL-based method adopted in the measurement of prompt D mesons. In comparison to previously published results based on the same data sample~\cite{Acharya:2019mgn}, the cross sections of prompt $\Dplus$ and $\Ds$ mesons have total uncertainties reduced by a factor ranging from 1.05 to 1.60 and cover an extended transverse-momentum range, down to $\pt=0$ and $\pt=1~\GeV/c$ for $\Dplus$ and $\Ds$ mesons, respectively. The measurements of non-prompt mesons were performed in the interval $1<\pt<24~\GeV/c$ for $\Dzero$ mesons, $2<\pt<16~\GeV/c$ for $\Dplus$ mesons, and $2<\pt<12~\GeV/c$ for $\Ds$ mesons. The measured $\pt$-differential cross sections are compatible with FONLL calculations in the full $\pt$ range of the measurements. For prompt D mesons, the measured values lie on the upper edge of the FONLL uncertainty band, while the measured non-prompt D-meson cross sections are in better agreement with the central value of the predictions obtained using the beauty-hadron cross section from FONLL calculations and the $\bhad\to\mathrm{D+X}$ decay kinematics from the PYTHIA~8 decayer. The GM-VFNS calculations also describe the measured prompt D-meson cross sections, while they underestimate 
the non-prompt D-meson cross sections. The modelling of the $\btoDX$ transition with a single step underestimates the measurements by a factor ranging between 2 and 10 depending on $\pt$. Larger cross sections, in better agreement with the data, are obtained with a two-step process in which the $\mathrm{b}\to\bhad$ fragmentation and the $\bhad\to\mathrm{D+X}$ decay kinematics are factorised.
Therefore, this does not invalidate the GM-VFNS calculation of the cross section of the partonic process, nor the validity of the collinear factorisation, but it confirms the importance of properly modelling the fragmentation process and the decay kinematics.

The ratios of production cross sections as well as the fragmentation fraction to strange mesons divided by the one to non-strange mesons for charm quarks,
\begin{equation*}
\bigg{(}\frac{\f{s}}{\f{u}+\f{d}}\bigg{)}_\mathrm{charm} = 0.136 \pm 0.005(\mathrm{stat}) \pm0.008(\mathrm{tot.~syst}),
\end{equation*}
and beauty quarks,
\begin{equation*}
\bigg{(}\frac{\f{s}}{\f{u}+\f{d}}\bigg{)}_\mathrm{beauty} = 0.127 \pm 0.036(\mathrm{stat})\pm0.014(\mathrm{tot.~syst}),
\end{equation*}
are compatible with previous measurements by other experiments for different centre-of-mass energies and colliding systems.

The $\bbbar$ production cross section at midrapidity per unit of rapidity in pp collisions at $\s=5.02~\TeV$ was estimated from the measured production cross sections of non-prompt $\Dzero$, $\Dplus$, and $\Ds$ mesons using the predictions based on FONLL calculations for the beauty-hadron cross section and the PYTHIA~8 decayer for the description of the $\bhad\to\mathrm{D+X}$ decay kinematics. The extrapolated $\de\sigma_\bbbar/\de y$ from each D-meson species are compatible among each other and with previous ALICE measurements based on dielectrons~\cite{Acharya:2020rfm}, and with FONLL and NNLO calculations. The $\de\sigma_\bbbar/\de y$ determined from the average of the three D-meson species is
\begin{equation*}
    \frac{\de\sigma_\bbbar}{\de y}\bigg{|}_{|y|<0.5} = 34.5 \pm 2.4 (\mathrm{stat}) ^{+4.7}_{-2.9} (\mathrm{tot.~syst})~\mub.
\end{equation*}

The measurements presented in this paper provide an important test for pQCD calculations in the charm and beauty sectors and a precise reference for studies in heavy-ion collisions.


\newenvironment{acknowledgement}{\relax}{\relax}
\begin{acknowledgement}
\section*{Acknowledgements}

The ALICE Collaboration would like to thank all its engineers and technicians for their invaluable contributions to the construction of the experiment and the CERN accelerator teams for the outstanding performance of the LHC complex.
The ALICE Collaboration gratefully acknowledges the resources and support provided by all Grid centres and the Worldwide LHC Computing Grid (WLCG) collaboration.
The ALICE Collaboration acknowledges the following funding agencies for their support in building and running the ALICE detector:
A. I. Alikhanyan National Science Laboratory (Yerevan Physics Institute) Foundation (ANSL), State Committee of Science and World Federation of Scientists (WFS), Armenia;
Austrian Academy of Sciences, Austrian Science Fund (FWF): [M 2467-N36] and Nationalstiftung f\"{u}r Forschung, Technologie und Entwicklung, Austria;
Ministry of Communications and High Technologies, National Nuclear Research Center, Azerbaijan;
Conselho Nacional de Desenvolvimento Cient\'{\i}fico e Tecnol\'{o}gico (CNPq), Financiadora de Estudos e Projetos (Finep), Funda\c{c}\~{a}o de Amparo \`{a} Pesquisa do Estado de S\~{a}o Paulo (FAPESP) and Universidade Federal do Rio Grande do Sul (UFRGS), Brazil;
Ministry of Education of China (MOEC) , Ministry of Science \& Technology of China (MSTC) and National Natural Science Foundation of China (NSFC), China;
Ministry of Science and Education and Croatian Science Foundation, Croatia;
Centro de Aplicaciones Tecnol\'{o}gicas y Desarrollo Nuclear (CEADEN), Cubaenerg\'{\i}a, Cuba;
Ministry of Education, Youth and Sports of the Czech Republic, Czech Republic;
The Danish Council for Independent Research | Natural Sciences, the VILLUM FONDEN and Danish National Research Foundation (DNRF), Denmark;
Helsinki Institute of Physics (HIP), Finland;
Commissariat \`{a} l'Energie Atomique (CEA) and Institut National de Physique Nucl\'{e}aire et de Physique des Particules (IN2P3) and Centre National de la Recherche Scientifique (CNRS), France;
Bundesministerium f\"{u}r Bildung und Forschung (BMBF) and GSI Helmholtzzentrum f\"{u}r Schwerionenforschung GmbH, Germany;
General Secretariat for Research and Technology, Ministry of Education, Research and Religions, Greece;
National Research, Development and Innovation Office, Hungary;
Department of Atomic Energy Government of India (DAE), Department of Science and Technology, Government of India (DST), University Grants Commission, Government of India (UGC) and Council of Scientific and Industrial Research (CSIR), India;
Indonesian Institute of Science, Indonesia;
Istituto Nazionale di Fisica Nucleare (INFN), Italy;
Institute for Innovative Science and Technology , Nagasaki Institute of Applied Science (IIST), Japanese Ministry of Education, Culture, Sports, Science and Technology (MEXT) and Japan Society for the Promotion of Science (JSPS) KAKENHI, Japan;
Consejo Nacional de Ciencia (CONACYT) y Tecnolog\'{i}a, through Fondo de Cooperaci\'{o}n Internacional en Ciencia y Tecnolog\'{i}a (FONCICYT) and Direcci\'{o}n General de Asuntos del Personal Academico (DGAPA), Mexico;
Nederlandse Organisatie voor Wetenschappelijk Onderzoek (NWO), Netherlands;
The Research Council of Norway, Norway;
Commission on Science and Technology for Sustainable Development in the South (COMSATS), Pakistan;
Pontificia Universidad Cat\'{o}lica del Per\'{u}, Peru;
Ministry of Education and Science, National Science Centre and WUT ID-UB, Poland;
Korea Institute of Science and Technology Information and National Research Foundation of Korea (NRF), Republic of Korea;
Ministry of Education and Scientific Research, Institute of Atomic Physics and Ministry of Research and Innovation and Institute of Atomic Physics, Romania;
Joint Institute for Nuclear Research (JINR), Ministry of Education and Science of the Russian Federation, National Research Centre Kurchatov Institute, Russian Science Foundation and Russian Foundation for Basic Research, Russia;
Ministry of Education, Science, Research and Sport of the Slovak Republic, Slovakia;
National Research Foundation of South Africa, South Africa;
Swedish Research Council (VR) and Knut \& Alice Wallenberg Foundation (KAW), Sweden;
European Organization for Nuclear Research, Switzerland;
Suranaree University of Technology (SUT), National Science and Technology Development Agency (NSDTA) and Office of the Higher Education Commission under NRU project of Thailand, Thailand;
Turkish Atomic Energy Agency (TAEK), Turkey;
National Academy of  Sciences of Ukraine, Ukraine;
Science and Technology Facilities Council (STFC), United Kingdom;
National Science Foundation of the United States of America (NSF) and United States Department of Energy, Office of Nuclear Physics (DOE NP), United States of America.
\end{acknowledgement}

\bibliographystyle{utphys}   
\bibliography{bibliography}

\newpage
\appendix

%
%

\section{The ALICE Collaboration}
\label{app:collab}
%
\begingroup
\small
\begin{flushleft}
S.~Acharya$^{\rm 142}$, 
D.~Adamov\'{a}$^{\rm 97}$, 
A.~Adler$^{\rm 75}$, 
J.~Adolfsson$^{\rm 82}$, 
G.~Aglieri Rinella$^{\rm 35}$, 
M.~Agnello$^{\rm 31}$, 
N.~Agrawal$^{\rm 55}$, 
Z.~Ahammed$^{\rm 142}$, 
S.~Ahmad$^{\rm 16}$, 
S.U.~Ahn$^{\rm 77}$
% ...

\section*{Affiliation notes}

$^{\rm I}$ Deceased\\
$^{\rm II}$ Also at: Italian National Agency for New Technologies, Energy and Sustainable Economic Development (ENEA), Bologna, Italy\\
$^{\rm III}$ Also at: Dipartimento DET del Politecnico di Torino, Turin, Italy\\
$^{\rm IV}$ Also at: M.V. Lomonosov Moscow State University, D.V. Skobeltsyn Institute of Nuclear, Physics, Moscow, Russia\\
$^{\rm V}$ Also at: Institute of Theoretical Physics, University of Wroclaw, Poland\\

\section*{Collaboration Institutes}

$^{1}$ A.I. Alikhanyan National Science Laboratory (Yerevan Physics Institute) Foundation, Yerevan, Armenia\\
$^{2}$ AGH University of Science and Technology, Cracow, Poland\\
$^{3}$ Bogolyubov Institute for Theoretical Physics, National Academy of Sciences of Ukraine, Kiev, Ukraine\\
$^{4}$ Bose Institute, Department of Physics  and Centre for Astroparticle Physics and Space Science (CAPSS), Kolkata, India\\
$^{5}$ Budker Institute for Nuclear Physics, Novosibirsk, Russia\\
$^{6}$ California Polytechnic State University, San Luis Obispo, California, United States\\
$^{7}$ Central China Normal University, Wuhan, China\\
$^{8}$ Centro de Aplicaciones Tecnol\'{o}gicas y Desarrollo Nuclear (CEADEN), Havana, Cuba\\
$^{9}$ Centro de Investigaci\'{o}n y de Estudios Avanzados (CINVESTAV), Mexico City and M\'{e}rida, Mexico\\
$^{10}$ Chicago State University, Chicago, Illinois, United States\\
$^{11}$ China Institute of Atomic Energy, Beijing, China\\
$^{12}$ Chungbuk National University, Cheongju, Republic of Korea\\
$^{13}$ Comenius University Bratislava, Faculty of Mathematics, Physics and Informatics, Bratislava, Slovakia\\
$^{14}$ COMSATS University Islamabad, Islamabad, Pakistan\\
$^{15}$ Creighton University, Omaha, Nebraska, United States\\
$^{16}$ Department of Physics, Aligarh Muslim University, Aligarh, India\\
$^{17}$ Department of Physics, Pusan National University, Pusan, Republic of Korea\\
$^{18}$ Department of Physics, Sejong University, Seoul, Republic of Korea\\
$^{19}$ Department of Physics, University of California, Berkeley, California, United States\\
$^{20}$ Department of Physics, University of Oslo, Oslo, Norway\\
$^{21}$ Department of Physics and Technology, University of Bergen, Bergen, Norway\\
$^{22}$ Dipartimento di Fisica dell'Universit\`{a} 'La Sapienza' and Sezione INFN, Rome, Italy\\
$^{23}$ Dipartimento di Fisica dell'Universit\`{a} and Sezione INFN, Cagliari, Italy\\
$^{24}$ Dipartimento di Fisica dell'Universit\`{a} and Sezione INFN, Trieste, Italy\\
$^{25}$ Dipartimento di Fisica dell'Universit\`{a} and Sezione INFN, Turin, Italy\\
$^{26}$ Dipartimento di Fisica e Astronomia dell'Universit\`{a} and Sezione INFN, Bologna, Italy\\
$^{27}$ Dipartimento di Fisica e Astronomia dell'Universit\`{a} and Sezione INFN, Catania, Italy\\
$^{28}$ Dipartimento di Fisica e Astronomia dell'Universit\`{a} and Sezione INFN, Padova, Italy\\
$^{29}$ Dipartimento di Fisica e Nucleare e Teorica, Universit\`{a} di Pavia  and Sezione INFN, Pavia, Italy\\
$^{30}$ Dipartimento di Fisica `E.R.~Caianiello' dell'Universit\`{a} and Gruppo Collegato INFN, Salerno, Italy\\
$^{31}$ Dipartimento DISAT del Politecnico and Sezione INFN, Turin, Italy\\
$^{32}$ Dipartimento di Scienze e Innovazione Tecnologica dell'Universit\`{a} del Piemonte Orientale and INFN Sezione di Torino, Alessandria, Italy\\
$^{33}$ Dipartimento di Scienze MIFT, Universit\`{a} di Messina, Messina, Italy\\
$^{34}$ Dipartimento Interateneo di Fisica `M.~Merlin' and Sezione INFN, Bari, Italy\\
$^{35}$ European Organization for Nuclear Research (CERN), Geneva, Switzerland\\
$^{36}$ Faculty of Electrical Engineering, Mechanical Engineering and Naval Architecture, University of Split, Split, Croatia\\
$^{37}$ Faculty of Engineering and Science, Western Norway University of Applied Sciences, Bergen, Norway\\
$^{38}$ Faculty of Nuclear Sciences and Physical Engineering, Czech Technical University in Prague, Prague, Czech Republic\\
$^{39}$ Faculty of Science, P.J.~\v{S}af\'{a}rik University, Ko\v{s}ice, Slovakia\\
$^{40}$ Frankfurt Institute for Advanced Studies, Johann Wolfgang Goethe-Universit\"{a}t Frankfurt, Frankfurt, Germany\\
$^{41}$ Fudan University, Shanghai, China\\
$^{42}$ Gangneung-Wonju National University, Gangneung, Republic of Korea\\
$^{43}$ Gauhati University, Department of Physics, Guwahati, India\\
$^{44}$ Helmholtz-Institut f\"{u}r Strahlen- und Kernphysik, Rheinische Friedrich-Wilhelms-Universit\"{a}t Bonn, Bonn, Germany\\
$^{45}$ Helsinki Institute of Physics (HIP), Helsinki, Finland\\
$^{46}$ High Energy Physics Group,  Universidad Aut\'{o}noma de Puebla, Puebla, Mexico\\
$^{47}$ Hiroshima University, Hiroshima, Japan\\
$^{48}$ Hochschule Worms, Zentrum  f\"{u}r Technologietransfer und Telekommunikation (ZTT), Worms, Germany\\
$^{49}$ Horia Hulubei National Institute of Physics and Nuclear Engineering, Bucharest, Romania\\
$^{50}$ Indian Institute of Technology Bombay (IIT), Mumbai, India\\
$^{51}$ Indian Institute of Technology Indore, Indore, India\\
$^{52}$ Indonesian Institute of Sciences, Jakarta, Indonesia\\
$^{53}$ INFN, Laboratori Nazionali di Frascati, Frascati, Italy\\
$^{54}$ INFN, Sezione di Bari, Bari, Italy\\
$^{55}$ INFN, Sezione di Bologna, Bologna, Italy\\
$^{56}$ INFN, Sezione di Cagliari, Cagliari, Italy\\
$^{57}$ INFN, Sezione di Catania, Catania, Italy\\
$^{58}$ INFN, Sezione di Padova, Padova, Italy\\
$^{59}$ INFN, Sezione di Roma, Rome, Italy\\
$^{60}$ INFN, Sezione di Torino, Turin, Italy\\
$^{61}$ INFN, Sezione di Trieste, Trieste, Italy\\
$^{62}$ Inha University, Incheon, Republic of Korea\\
$^{63}$ Institute for Gravitational and Subatomic Physics (GRASP), Utrecht University/Nikhef, Utrecht, Netherlands\\
$^{64}$ Institute for Nuclear Research, Academy of Sciences, Moscow, Russia\\
$^{65}$ Institute of Experimental Physics, Slovak Academy of Sciences, Ko\v{s}ice, Slovakia\\
$^{66}$ Institute of Physics, Homi Bhabha National Institute, Bhubaneswar, India\\
$^{67}$ Institute of Physics of the Czech Academy of Sciences, Prague, Czech Republic\\
$^{68}$ Institute of Space Science (ISS), Bucharest, Romania\\
$^{69}$ Institut f\"{u}r Kernphysik, Johann Wolfgang Goethe-Universit\"{a}t Frankfurt, Frankfurt, Germany\\
$^{70}$ Instituto de Ciencias Nucleares, Universidad Nacional Aut\'{o}noma de M\'{e}xico, Mexico City, Mexico\\
$^{71}$ Instituto de F\'{i}sica, Universidade Federal do Rio Grande do Sul (UFRGS), Porto Alegre, Brazil\\
$^{72}$ Instituto de F\'{\i}sica, Universidad Nacional Aut\'{o}noma de M\'{e}xico, Mexico City, Mexico\\
$^{73}$ iThemba LABS, National Research Foundation, Somerset West, South Africa\\
$^{74}$ Jeonbuk National University, Jeonju, Republic of Korea\\
$^{75}$ Johann-Wolfgang-Goethe Universit\"{a}t Frankfurt Institut f\"{u}r Informatik, Fachbereich Informatik und Mathematik, Frankfurt, Germany\\
$^{76}$ Joint Institute for Nuclear Research (JINR), Dubna, Russia\\
$^{77}$ Korea Institute of Science and Technology Information, Daejeon, Republic of Korea\\
$^{78}$ KTO Karatay University, Konya, Turkey\\
$^{79}$ Laboratoire de Physique des 2 Infinis, Ir\`{e}ne Joliot-Curie, Orsay, France\\
$^{80}$ Laboratoire de Physique Subatomique et de Cosmologie, Universit\'{e} Grenoble-Alpes, CNRS-IN2P3, Grenoble, France\\
$^{81}$ Lawrence Berkeley National Laboratory, Berkeley, California, United States\\
$^{82}$ Lund University Department of Physics, Division of Particle Physics, Lund, Sweden\\
$^{83}$ Moscow Institute for Physics and Technology, Moscow, Russia\\
$^{84}$ Nagasaki Institute of Applied Science, Nagasaki, Japan\\
$^{85}$ Nara Women{'}s University (NWU), Nara, Japan\\
$^{86}$ National and Kapodistrian University of Athens, School of Science, Department of Physics , Athens, Greece\\
$^{87}$ National Centre for Nuclear Research, Warsaw, Poland\\
$^{88}$ National Institute of Science Education and Research, Homi Bhabha National Institute, Jatni, India\\
$^{89}$ National Nuclear Research Center, Baku, Azerbaijan\\
$^{90}$ National Research Centre Kurchatov Institute, Moscow, Russia\\
$^{91}$ Niels Bohr Institute, University of Copenhagen, Copenhagen, Denmark\\
$^{92}$ Nikhef, National institute for subatomic physics, Amsterdam, Netherlands\\
$^{93}$ NRC Kurchatov Institute IHEP, Protvino, Russia\\
$^{94}$ NRC \guillemotleft Kurchatov\guillemotright  Institute - ITEP, Moscow, Russia\\
$^{95}$ NRNU Moscow Engineering Physics Institute, Moscow, Russia\\
$^{96}$ Nuclear Physics Group, STFC Daresbury Laboratory, Daresbury, United Kingdom\\
$^{97}$ Nuclear Physics Institute of the Czech Academy of Sciences, \v{R}e\v{z} u Prahy, Czech Republic\\
$^{98}$ Oak Ridge National Laboratory, Oak Ridge, Tennessee, United States\\
$^{99}$ Ohio State University, Columbus, Ohio, United States\\
$^{100}$ Petersburg Nuclear Physics Institute, Gatchina, Russia\\
$^{101}$ Physics department, Faculty of science, University of Zagreb, Zagreb, Croatia\\
$^{102}$ Physics Department, Panjab University, Chandigarh, India\\
$^{103}$ Physics Department, University of Jammu, Jammu, India\\
$^{104}$ Physics Department, University of Rajasthan, Jaipur, India\\
$^{105}$ Physikalisches Institut, Eberhard-Karls-Universit\"{a}t T\"{u}bingen, T\"{u}bingen, Germany\\
$^{106}$ Physikalisches Institut, Ruprecht-Karls-Universit\"{a}t Heidelberg, Heidelberg, Germany\\
$^{107}$ Physik Department, Technische Universit\"{a}t M\"{u}nchen, Munich, Germany\\
$^{108}$ Politecnico di Bari and Sezione INFN, Bari, Italy\\
$^{109}$ Research Division and ExtreMe Matter Institute EMMI, GSI Helmholtzzentrum f\"ur Schwerionenforschung GmbH, Darmstadt, Germany\\
$^{110}$ Russian Federal Nuclear Center (VNIIEF), Sarov, Russia\\
$^{111}$ Saha Institute of Nuclear Physics, Homi Bhabha National Institute, Kolkata, India\\
$^{112}$ School of Physics and Astronomy, University of Birmingham, Birmingham, United Kingdom\\
$^{113}$ Secci\'{o}n F\'{\i}sica, Departamento de Ciencias, Pontificia Universidad Cat\'{o}lica del Per\'{u}, Lima, Peru\\
$^{114}$ St. Petersburg State University, St. Petersburg, Russia\\
$^{115}$ Stefan Meyer Institut f\"{u}r Subatomare Physik (SMI), Vienna, Austria\\
$^{116}$ SUBATECH, IMT Atlantique, Universit\'{e} de Nantes, CNRS-IN2P3, Nantes, France\\
$^{117}$ Suranaree University of Technology, Nakhon Ratchasima, Thailand\\
$^{118}$ Technical University of Ko\v{s}ice, Ko\v{s}ice, Slovakia\\
$^{119}$ The Henryk Niewodniczanski Institute of Nuclear Physics, Polish Academy of Sciences, Cracow, Poland\\
$^{120}$ The University of Texas at Austin, Austin, Texas, United States\\
$^{121}$ Universidad Aut\'{o}noma de Sinaloa, Culiac\'{a}n, Mexico\\
$^{122}$ Universidade de S\~{a}o Paulo (USP), S\~{a}o Paulo, Brazil\\
$^{123}$ Universidade Estadual de Campinas (UNICAMP), Campinas, Brazil\\
$^{124}$ Universidade Federal do ABC, Santo Andre, Brazil\\
$^{125}$ University of Cape Town, Cape Town, South Africa\\
$^{126}$ University of Houston, Houston, Texas, United States\\
$^{127}$ University of Jyv\"{a}skyl\"{a}, Jyv\"{a}skyl\"{a}, Finland\\
$^{128}$ University of Kansas, Lawrence, Kansas, United States\\
$^{129}$ University of Liverpool, Liverpool, United Kingdom\\
$^{130}$ University of Science and Technology of China, Hefei, China\\
$^{131}$ University of South-Eastern Norway, Tonsberg, Norway\\
$^{132}$ University of Tennessee, Knoxville, Tennessee, United States\\
$^{133}$ University of the Witwatersrand, Johannesburg, South Africa\\
$^{134}$ University of Tokyo, Tokyo, Japan\\
$^{135}$ University of Tsukuba, Tsukuba, Japan\\
$^{136}$ Universit\'{e} Clermont Auvergne, CNRS/IN2P3, LPC, Clermont-Ferrand, France\\
$^{137}$ Universit\'{e} de Lyon, CNRS/IN2P3, Institut de Physique des 2 Infinis de Lyon , Lyon, France\\
$^{138}$ Universit\'{e} de Strasbourg, CNRS, IPHC UMR 7178, F-67000 Strasbourg, France, Strasbourg, France\\
$^{139}$ Universit\'{e} Paris-Saclay Centre d'Etudes de Saclay (CEA), IRFU, D\'{e}partment de Physique Nucl\'{e}aire (DPhN), Saclay, France\\
$^{140}$ Universit\`{a} degli Studi di Foggia, Foggia, Italy\\
$^{141}$ Universit\`{a} di Brescia and Sezione INFN, Brescia, Italy\\
$^{142}$ Variable Energy Cyclotron Centre, Homi Bhabha National Institute, Kolkata, India\\
$^{143}$ Warsaw University of Technology, Warsaw, Poland\\
$^{144}$ Wayne State University, Detroit, Michigan, United States\\
$^{145}$ Westf\"{a}lische Wilhelms-Universit\"{a}t M\"{u}nster, Institut f\"{u}r Kernphysik, M\"{u}nster, Germany\\
$^{146}$ Wigner Research Centre for Physics, Budapest, Hungary\\
$^{147}$ Yale University, New Haven, Connecticut, United States\\
$^{148}$ Yonsei University, Seoul, Republic of Korea\\

\bigskip 

\end{flushleft} 
\endgroup  
\end{document}